\documentclass[aps,supepscriptaddress,showpacs,nofootinbib]{revtex4-1}
\usepackage{amsmath}
\usepackage{bm}
\usepackage{graphicx}
\usepackage{tikz}
\usepackage{bm}
\usepackage{chemarrow}
\usepackage{float}
\usepackage{color}
\usepackage{soul}

\begin{document}

\title{Extended Thermodynamic and Mechanical Evolution Criterion for Fluids}

\author{David Hochberg}

\author{Isabel Herreros}
\email{Corresponding author: iherreros@cab.inta-csic.es} 
\altaffiliation[also at:
]{Departamento de Ingenier\'{i}a T\'{e}rmica y Fluidos, Universidad
Carlos III de Madrid, 28911 Legan\'{e}s, Madrid, Spain}
\affiliation{Centro de
Astrobiolog\'{\i}a (CAB) CSIC-INTA, Carretera Ajalvir Kil\'{o}metro
4, 28850 Torrej\'{o}n de Ardoz, Madrid, Spain}

\begin{abstract}
The Glansdorff and Prigogine General Evolution Criterion (GEC) is an
inequality that holds for macroscopic physical systems obeying local
equilibrium and that are constrained under time-independent boundary conditions. 
The latter, however, may prove overly restrictive for many applications
involving fluid flow in physics, chemistry and biology. We therefore analyze in detail a physically more-encompassing
evolution criterion for time-dependent convective viscous flow with
time-dependent boundary conditions. The result is an inequality involving the sum of a bulk volume and a surface contribution, and reduces to the GEC if and only
if the surface term is zero. We use the closed-form analytical
solution of the starting flow problem in straight cylindrical pipes to confirm this extended general evolution
criterion. We next validate the
starting flow problem and evolution criterion numerically. Numerical methods are used to
validate the extended evolution criterion in non-fully developed
flows inside complex geometries with curvature and torsion, such as encountered in sections of helical pipes.
Knowledge of only the algebraic sign of the surface contribution is sufficient for predicting how the volume thermodynamic forces evolve in time and how the system
approaches its non-equilibrium stationary state consistent with the boundary conditions.  
\end{abstract}


\maketitle

\section{\label{sec:intro} Introduction}

A unifying non-equilibrium thermodynamic principle for dissipative systems including mechanical processes was established many years ago by Glansdorff and Prigogine. They derived a general inequality involving the generation or production of entropy valid for the entire range of macroscopic physics. Their result states that the change in the thermodynamic forces proceeds always in a way such as to lower the entropy production. This is known as the General Evolution Criterion (GEC) \cite{GP1964}. As early as 1962, and prior to establishing the GEC two years later, those authors had applied the principles of non-equilibrium thermodynamics to viscous fluid flows and restricted to systems close to equilibrium \cite{G&P-fluids}.  Henceforth, the non-equilibrium thermodynamic principles established by them have attained enduring relevance across various scientific domains. Such principles provide essential insights into the behavior of dissipative systems, their connection with life's emergence, and their application in understanding complex phenomena such as fluid flows.

A crucial point in the original proof of the GEC is that it relies on the assumption of time-independent, or fixed boundary conditions \cite{GP1964}. When considering a finite volume system with a corresponding bounding surface, the demonstration of the GEC requires that the chemical potentials $\mu_{\gamma}$, the temperature $T$, and any external mechanical forces $\bm{F}$ acting on this boundary do not depend on the time.  In the case of convective processes, the fluid velocity on this surface $\bm{v}$ also cannot depend on the time.  Nevertheless, these fixed boundary conditions can all prove to be excessively restrictive for describing real systems. A most noteworthy case in point are living organisms, viewed as ordered dissipative structures, which sustain their existence and behaviors through a continuous exchange of energy and matter with their environment. They are neither isolated nor closed systems, but are open systems 
intimately linked to their surroundings, precisely via mass and energy transport across their boundaries. 

Time-dependent boundary conditions play a major role over a wide range of scientific disciplines. Their usefulness lies in the capacity to model and decipher dynamic systems, capturing the ever-changing nature of real-world phenomena. By integrating these conditions into their analyses, scientists can furnish more precise predictions and design better solutions, which, in turn, drives our understanding of intricate natural processes and fosters advancements in technology and science. The importance of open systems subject to boundary conditions is of course solidly established in physics, engineering and applied mathematics. By contrast, in chemistry, the GEC has rarely been used, if at all, to justify experimental results, and most likely because its application is severely limited to systems with fixed boundary conditions.  Only very recently has a thermodynamic evolution theorem been derived for reacting chemical systems subject to the open flow of matter \cite{Hochberg2020}.  In the field of biology, basic fundamental physiological phenomena in living systems can be considered to be consequences of the fact that organisms are quasi-stationary open systems \cite{Bertalanffy1950}. A striking example of openness in biology is the fact that cellular processes are guided and constrained by imposed external boundary conditions \cite{Vahey2014}.  In a closely related context, the rich repertoire of interactions between living cells, hydrodynamic stresses, and fluid flow-induced effects imposed on cells is reviewed in \cite{Huber2018}. 

There are therefore compelling reasons calling for a more encompassing evolution criterion, one that goes beyond the original GEC, one which includes time-dependent boundary conditions for understanding the thermodynamic and mechanical evolution of open systems. Although a general inequality allowing for time-dependent boundary conditions was derived in the early works by these authors \cite{GP1964, GPbook}, to the best of our knowledge, it has not been used, developed, nor applied to any specific non-equilibrium macroscopic physical system. Since convective fluid flow is of particular interest, importance and relevance over a wide range of problems in both applied and pure science, we consider in detail an evolution criterion for time dependent convective processes, where the velocity field on the system’s boundary can have any time dependence. We analyze an inequality for an expression having physical units of entropy production. It involves a surface and bulk volume contribution, the sum of both are constrained by this inequality. The bulk term is the product of currents times the rate of change of the thermodynamic forces that produce those currents. The implications of this extended inequality are far-reaching. Indeed, once we recognize that the boundary conditions can be chosen freely, that they correspond to the thermodynamic/mechanical state of the environment external to the system, then we conclude that the state of the environment affects the internal thermodynamic/mechanical evolution of the system. Thus, external influences, as codified through the boundary conditions, affect the way the system’s thermodynamic forces evolve in time.   We provide specific examples of this where, by simple choice of the fluid velocity on the entrance and exit bounding surfaces of both straight and curved pipes, we can either make the internal fluid viscous entropy production increase to a maximum value, or else decrease to a minimum.  

This paper is organized as follows.  In Sec \ref{sec:GECtime} we introduce the local form of the joint mechanical-thermodynamic inequality that holds for incompressible newtonian hydrodynamics and discuss its physical interpretation.  Integration of this local inequality over the system volume yields a surface term plus a bulk volume contribution which highlight the interplay between the mechanical forces and flows acting at the system boundary and the generalized thermodynamic forces evolving within the system’s interior. 
We identify the explicit form of the thermodynamic forces and flows and the expressions for the corresponding surface and volume integrals used in this paper. A brief resume of the rate of strain tensor, the velocity field and viscous entropy production for Poiseuille flow, is given in Sec \ref {sec:Poiseuille}, which serve as important benchmarks for detailed comparison with the starting flow problem in a pipe, the analytical solution of which is exploited in Sec \ref{sec:startPoiseuille}. We evaluate the time-dependent viscous entropy production and prove that it increases from zero to a final stationary positive value for all positive values of the pipe radius, remaining null only on the pipe's central axis. We use the analytical solution of the starting flow problem to evaluate the surface and volume integrals that appear in the extended evolution inequality.  The volume contribution yields a positive value for all times, and tends to zero from above asymptotically. This shows that the GEC does not hold in this problem involving time-dependent boundary conditions. The surface contribution is negative for all times, tending to zero asymptotically from below.  There is a partial cancellation between surface and volume contributions yielding a finite remainder for the latter which is negative definite and goes to zero from below as the system approaches its NESS.  This proves that the starting flow problem obeys the extended GEC (EGEC) presented herein.  

We then test the EGEC for more complex, curved pipe geometries and boundary conditions, where non-fully developed flows are expected, and thus requiring the use of numerical approximations.  In Sec \ref{sec:Numerical_approximation} we summarize the mathematical and numerical models and present the discretized expressions for the surface and volume contributions which follow from applying the Finite Element Method (FEM). We validate our FEM model by first solving the starting flow problem for a Poiseuille flow using this numerical procedure. Comparison of the boundary conditions for the fluid velocity at entrance and exit surfaces with an intermediate sectional control surface proves that the numerical solution corresponds to a fully developed velocity, in complete accord with the analytical expression. Evolution of the velocity on the central axis for entrance, exit and control surfaces also supports this claim. Numerical evaluation of the individual surface, volume and the combined surface plus volume contributions reveal an excellent agreement with the analytical results presented in Sec \ref{sec:startPoiseuille}.  With our numerical approximation thus validated, we apply it to helical geometries in which secondary flows are also generated.  The EGEC for helical pipe flows is treated in Sec \ref{sec:helical} for two contrasting cases: (i) for fixed boundary conditions and (ii) for time-dependent boundary conditions. In the former the calculation of the axial component of the velocity at inlet, outlet and two interior control surfaces of the helical conduit demonstrates that the flow is not fully developed. The secondary velocity also reveals the existence of a single circulating vortex, not fully developed, also in accord with our previous studies on flow in curved pipes with torsion.  Computation of the surface integral yields a null result (due to fixed boundary conditions) whereas the volume integral implies a negative contribution that tends to zero from below as the system approaches its NESS. The GEC is therefore obeyed. When we impose time-dependent boundary conditions, the calculated axial and secondary velocity components still confirm the presence of non-fully developed flows. Now by contrast, the surface integral returns time-dependent negative values, while the volume integral is positive definite. Nevertheless the sum of these two contributions obeys the EGEC.
Discussion and concluding remarks are given in Sec \ref{sec:disc}.

\section{\label{sec:GECtime} Evolution criterion for time-dependent convection processes}

The General Evolution Criterion (GEC) encompasses time-dependent
convective processes in macroscopic physics, the details of its
derivation can be found in \cite{GP1964} and \cite{GPbook}. Those
considerations began by deriving explicit expressions for the second
order differential of the specific entropy $\delta^2 s$. The entropy
plays a privileged role in the thermodynamic theory of
stability. This is because, for an isolated system, entropy $s$ must
be a maximum at equilibrium. Thus, the first order differential is
zero $\delta s = 0$ and the second order differential $\delta^2 s <
0$ must therefore be \textit{negative}. The development of explicit
expressions for the second-order variation $\delta^2 s$ is based on
using the classical Gibbs formula together with the conservation, or
balance, equations for mass, momentum and energy. By invoking the
\textit{local equilibrium hypothesis}, one arrives at a negative
semi-definite form: $\Psi \leq 0$ which is related to $\delta^2 s
\leq 0$, and which is valid for non-equilibrium macroscopic physical
systems. In this paper, we are interested in the application and
interpretation of the extended evolution criterion (EGEC) to viscous fluid flow. In
the case of \textit{time-dependent} convective processes (with no transport of heat or matter), the
explicit expression for the local quantity $\Psi$, can be simplified
and written as follows (see (9.77) in \cite{GPbook} and (3.24) in
\cite{GP1964}):
\begin{equation}\label{GECtime}
\boxed{\Psi = [P_{ij}T^{-1}(\partial_t \bm{v}_i)],_j -(P_{ij} + \rho
\bm{v}_i\bm{v}_j)\, \partial_t(T^{-1}\bm{v}_i),_j + \rho\bm{v}_j\,
\partial_t \big(T^{-1} \frac{1}{2} \bm{v}^2 \big),_j \leq 0.}
\end{equation}

\begin{figure}[htb]
\begin{center}
\includegraphics[width=0.70\textwidth]{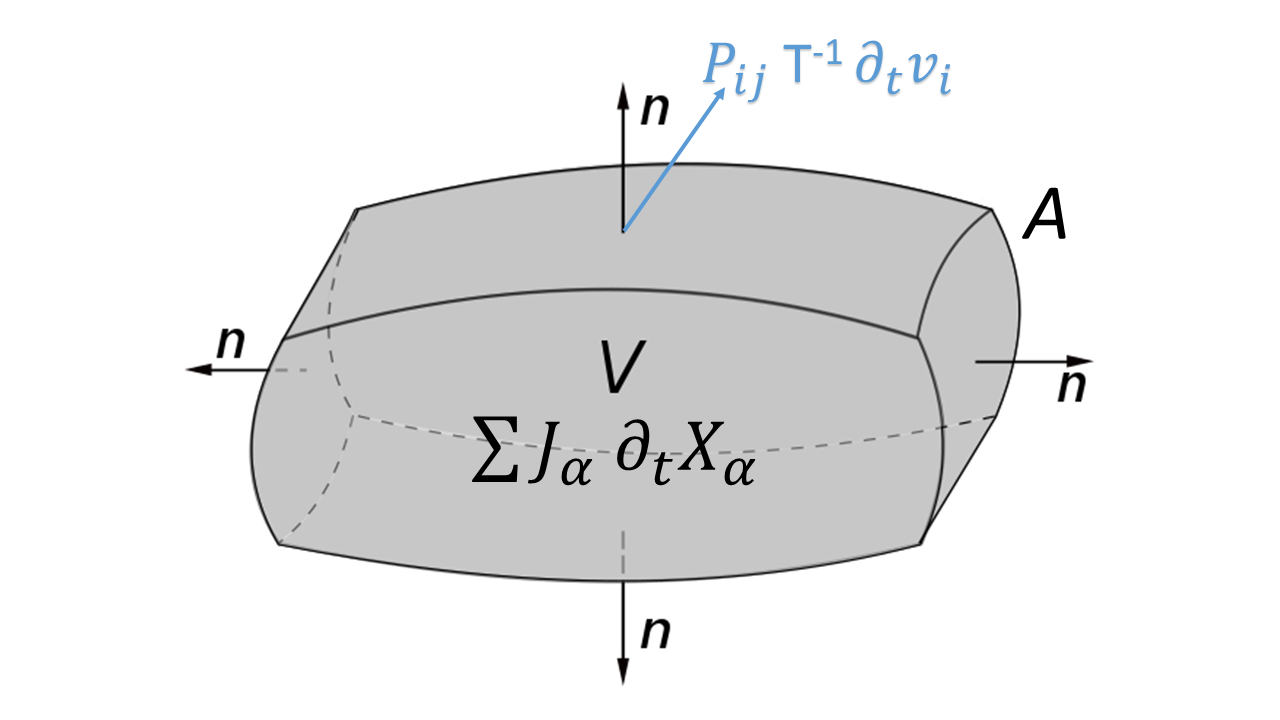}
\end{center}
\caption{\label{balanceeqn} System volume $V$ with bounding surface
$A$ showing the bulk source term and the flux term contributions to
Eq. (\ref{weakGEC}). The GEC follows for time independent boundary
conditions: $\partial_t v_i|_{A = \partial V} = 0$. Otherwise, a
more general evolution criterion holds which involves both surface
and volume contributions, Eq.(\ref{weakGEC}). }
\end{figure}

For constant temperature $T$ and fluid density $\rho$, and pressure
$p$, $\Psi$ is a function of the spatial and temporal derivatives of
the fluid velocity $\bm{v}$. Here $P_{ij} = p \, \delta_{ij} +
p_{ij}$, the term $-p_{ij}$ is the deviatoric stress tensor, see below and \cite{GPbook}. The
boxed expression is the local statement of a much more general
evolution criterion than the GEC, and for purely time-dependent
fluid motion. We note the above expression contains a flow
term (first term on the rhs) and a source term (the second
and third terms) related to the evolution inside the system. The
first contribution in Eq.(\ref{GECtime}) is the divergence of a flux,
or current. The validation and interpretation of Eq. (\ref{GECtime})
for fluid flow in both straight and curved conduits is one of the
major goals of this paper. $\Psi$ is zero only in
non-equilibrium stationary states (NESS), as well as for
equilibrium. By integrating $\Psi$ over the system volume $V$, we
obtain a surface integral and a bulk volume term:
\begin{equation}\label{weakGEC}
\int_V \Psi \, dV = \int_{A = \partial V} [P_{ij}T^{-1}(\partial_t
\bm{v}_i)]{\hat n}_j \, dA + \int_V \sum_{\alpha} J_{\alpha}\,\,
\partial_t X_{\alpha}\,\, dV\leq 0.
\end{equation}
The sum of surface and volume terms must be negative semi-definite.
The outward unit normal to the bounding surface $A$ is denoted by
${\hat n}$, see Fig. \ref{balanceeqn}. The definition of the
generalized thermodynamic flows $J_{\alpha}$ and forces $X_{\alpha}$
appearing in the volume integral term can be read off directly from
comparing to the two source terms in Eq. (\ref{GECtime}), and are
listed individually in Table \ref{JXtimedependent}. Note, if we
define $\Phi$ such that $\frac{\partial \Phi}{\partial t} =\int_V
\Psi \, dV$, then Eq. (\ref{weakGEC}) has the structure of a
balance equation. The source term is given by $\sum_{\alpha} J_{\alpha}\,
\partial_t X_{\alpha}$ whereas the current or flux entering and
exiting the system is given by $j_j = P_{ij}T^{-1}
\partial_t \bm{v}_i$, see Fig \ref{balanceeqn}.
If the time-independent (fixed) boundary conditions invoked in
\cite{GPbook} (constant temperature $T$ and chemical potentials
$\mu_{\gamma}$ on the bounding surface $A$) are extended to include
the fluid velocity on the surface
\begin{equation}\label{vcondition2}
[\partial_t (\bm{v}_i)]_A = 0,
\end{equation}
then the surface integral in Eq. (\ref{weakGEC}) vanishes
identically, and in \textit{this} case, and only in this case, do we
obtain the General Evolution Criterion (GEC) \cite{GP1964} as it applies to fluid
dynamics:
\begin{equation}\label{GECglobal}
\frac{\partial_{X} \Phi}{\partial t} \equiv \int_V \sum_{\alpha}
J_{\alpha}\,
\partial_t X_{\alpha}\,\, dV\leq 0 \qquad (\rm GEC).
\end{equation}
During the evolution of the macroscopic fluid system, this quantity
(\ref{GECglobal}) is negative and vanishes when the system reaches a
(non-equilibrium) steady state consistent with the \textit{fixed}
boundary conditions Eq. (\ref{vcondition2}). Physically, this means
that the temporal change in the generalized thermodynamic forces
$\partial_t X_{\alpha}$ proceeds always in a way as the lower the
value of the entropy production \cite{GPbook}, as defined by the
bilinear form in Eq. (\ref{source}) below.

But as explained in the Introducion, time independent boundary
conditions are very restrictive. Depending on the fluid mechanical
problem at hand, we may need to allow for time-dependent fluid
velocities on the surface $A$ bounding the volume, as for example,
specified input and output velocities, so the surface integral in
Eq. (\ref{weakGEC}) would not vanish in such cases, see Fig
\ref{balanceeqn}. Or free boundary conditions, in which case the
velocity on the system surface is not specified \textit{a-priori}.
We consider this generalization below (see Sec
\ref{sec:startPoiseuille}). Thus, the \textit{joint} evolution of
thermodynamic forces $X_{\alpha}$ within the system and the fluid
velocity on the boundary $[\partial_t (\bm{v}_i)]_A$ must be such as
to keep the quantity $\Psi$ negative, reaching zero only for
stationary states. So, we can state that the GEC can only hold when
there are no flows or currents at the boundary, a rather restrictive
situation. Otherwise, a more general evolution criterion results:
Eq. (\ref{weakGEC}), which necessarily involves the boundary
conditions.
\begin{table}[!tbh]
\begin{tabular}{lll}
\toprule Process & Flows $J_{\alpha}$\, & Forces $X_{\alpha}$\\
\hline Fluid mechanics &
$[P_{ij} + \rho v_i v_j]$ & \,\, $(-T^{-1}v_{i})_{,j}$ \\
Fluid mechanics & $\rho v_j$ & \,\, $(T^{-1}\frac{v^2}{2})_{,j}$ \\
\botrule
\end{tabular}
\caption{\label{JXtimedependent} Generalized forces $X_{\alpha}$ and
flows $J_{\alpha}$ for fluid dynamics. See Eqs.
(\ref{GECtime},\ref{weakGEC}).}
\end{table}

In the following, we work out the explicit expressions for the
surface and volume integrals that are needed to calculate Eq.
(\ref{weakGEC}).

\subsection{\label{sec:boundary} Surface integral of flow term for $\Psi$}

In realistic fluid flow problems, we may need to pay careful
attention to the boundary conditions, and so revisit the assumption
Eq. (\ref{vcondition2}). Thus,  integrating the first term in $\Psi$
over the volume yields (see Eq.(\ref{weakGEC}))
\begin{eqnarray}\label{surface}
\int_A \big(P_{ij} T^{-1} \partial_t v_i \big)n_j \, dA  &=&
\int_{A} \big[\big(p T^{-1} \partial_t v_j \big)n_j + \big(p_{ij} T^{-1}
\partial_t v_i \big)n_j\big] \,\, dA,
\end{eqnarray}
using $P_{ij} = p \, \delta_{ij} + p_{ij}$. Note: the deviatoric
stress tensor $d_{ij}= -p_{ij}$.
\subsection{\label{sec:fluidflow} Entropy production per unit volume}

Recall, the local source of entropy production is given by the
product of flows $J_{\alpha}$ and forces $X_{\alpha}$ \cite{GPbook}:
\begin{equation}\label{source}
\sigma[S] = \sum_{\alpha} J_{\alpha}\,\, X_{\alpha} \geq 0.
\end{equation}
Working this out explicitly, using Table \ref{JXtimedependent}, we
obtain
\begin{eqnarray}\label{source2}
\sum_{\alpha}\ J_{\alpha}\,\, X_{\alpha} &=& -[P_{ij} + \rho v_i
v_j] (T^{-1}v_{i})_{,j} + \rho v_j
(T^{-1}\frac{v^2}{2})_{,j}\\\label{source2b} &=& -T^{-1} P_{ij}
v_{i,j} = -T^{-1}\big(p \delta_{ij} +  p_{ij} \big)
v_{i,j}\\\label{source2c} &=& -T^{-1} p_{ij} v_{i,j} = + T^{-1}
d_{ij} v_{i,j}\\\label{visc}
&=& 2 \mu T^{-1} ||\bm{e}||^2 \geq 0\\
& \equiv& \sigma^{visc}[S].
\end{eqnarray}
We assume a uniform and steady temperature $T$. The second line
(\ref{source2b}) follows from the definition of $P_{ij}$, see
\cite{GPbook}; note in passing from first (\ref{source2}) to second
line (\ref{source2b}) the terms proportional to density $\rho$ have
canceled out. The third line (\ref{source2c}) follows from
incompressibility $v_{j,j} = 0$, so the dependence on the pressure
$p$ drops out, and here we use the fact that the deviatoric stress
tensor $d_{ij}= -p_{ij}$, and $d_{ij}= 2 \mu e_{ij}$. The fourth
line (\ref{visc}) is written in terms of the square modulus of the
rate-of strain tensor with components $e_{ij} =
\frac{1}{2}\Big(\frac{\partial v_i}{\partial x_j} + \frac{\partial
v_j}{\partial x_i}\Big)$, see \cite{HHPoF}. $\mu$ is the dynamic
viscosity.  So, Eq. (\ref{source}) yields the viscous contribution
to the local entropy production, or entropy source
$\sigma^{visc}[S]$, and is seen to be manifestly positive definite.

\subsection{\label{sec:GECreduction} Volume term for $\Psi$}

\bigskip

Using Table \ref{JXtimedependent}, after some simple algebra, we
have
\begin{eqnarray}\label{GECreduction0}
\sum_{\alpha} J_{\alpha}\, \partial_t X_{\alpha} &=& -(P_{ij} + \rho
\bm{v}_i\bm{v}_j)\,\, \partial_t(T^{-1}\bm{v}_i),_j +
\rho\bm{v}_j\,\,
\partial_t (T^{-1} \frac{1}{2} \bm{v}^2),_j\\
&=& - T^{-1} [p \delta_{ij} + p_{ij}]\, \partial_t(v_{i,j}) + \rho
T^{-1} v_{j} v_{k,j}\,
\partial_t(v_k)\\ \label{GECreduction2} &=& +T^{-1} d_{ij}\,
\partial_t(v_{i,j}) + \rho T^{-1} v_{j} v_{k,j}\, \partial_t(v_k).
\end{eqnarray}
Again, incompressibility implies the result does not depend on the
pressure $p$.

Note, we flip the sign in the intermediate step where we replaced
$-p_{ij}$ by $d_{ij}$.

\section{\label{sec:Poiseuille} Poiseuille flow}

We calculate the entropy production (per unit volume) due to
fully developed \textit{stationary} flow in a long cylindrical pipe
of constant circular cross section.  In the following section, we
then consider the evolution criterion Eq. (\ref{weakGEC}) using the
available analytical expression for the starting flow in such pipes.
This time-dependent flow approaches stationarity, converging rapidly
to Poiseuille flow on an appropriate time scale.

For Poiseuille flow, the only non-vanishing component of the
rate-of-strain tensor is
\begin{equation}\label{deviatoric}
e_{zr} = \frac{1}{2}\frac{\partial v_z(r)}{\partial r},
\end{equation}
where the axial or downstream velocity is a function of the radius
$(0 \leq r \leq a)$:
\begin{equation}
v_z(r) = \frac{G}{4\mu}\big(a^2 - r^2\big),
\end{equation}
where $\partial p/\partial z = -G$, with $G > 0$, is the pressure gradient. 

Substituting this expression into Eq.(\ref{visc}) yields the viscous
entropy production per unit volume:
\begin{equation}\label{PoisseuilleEntropy}
\sigma^{visc}[S] = \frac{G^2 r^2}{8 \mu T}
> 0,
\end{equation}
We can average this, for example, over the cross-sectional area of
the pipe:
\begin{equation}
<\sigma^{visc}[S]>_{area} = \frac{1}{\pi a^2}\int_0^a r\, dr
\int_0^{2\pi} d\phi \, \Big(\frac{G^2 r^2}{8 \mu T} \Big) =
\frac{G^2 a^2}{16 \mu T}.
\end{equation}
%

\section{\label{sec:startPoiseuille} Starting flow in a pipe}

It is useful to have a time-dependent analytic solution with which
to apply and validate the evolution criterion. To this end, we can
use the starting flow problem in a straight cylinder. We make use of
the known analytic time-dependent solution and use it to test the
GEC and its extension, as the system approaches its NESS: stationary
Poiseuille flow. In this case the entropy production
\textit{increases} from zero and tends to a final positive value as
the system reaches its NESS. In which case, the GEC as it stands
clearly cannot be obeyed (see details below), and we therefore must
assess the impact of the time-dependent boundary conditions.  The
starting flow problem is solved by Batchelor \cite{Batchelor} in
Sec. 4.3 and see Eq.(4.3.19), where the boundary and initial
conditions are also specified. Following his parametrization:
\begin{equation}\label{start}
v_z(r,t) = \frac{G}{4 \mu}\big(a^2 - r^2\big) -
\frac{2Ga^2}{\mu}\sum_{n=1}^{\infty} \frac{J_0(\lambda_n
\frac{r}{a})}{\lambda^3_n J_1(\lambda_n)} \exp \big(-\lambda^2_n
\frac{\nu t}{a^2} \big),
\end{equation}
so that
\begin{equation}\label{drstart}
\frac{\partial v_z(r,t)}{\partial r} = \frac{-Gr}{2\mu} +
\frac{2Ga}{\mu}\sum_{n=1}^{\infty} \frac{J_1(\lambda_n
\frac{r}{a})}{\lambda^2_n J_1(\lambda_n)} \exp \big(-\lambda^2_n
\frac{\nu t}{a^2} \big),
\end{equation}
and
\begin{equation}\label{dtstart}
\frac{\partial v_z(r,t)}{\partial t} =  +
\frac{2G}{\rho}\sum_{n=1}^{\infty} \frac{J_0(\lambda_n
\frac{r}{a})}{\lambda_n J_1(\lambda_n)} \exp \big(-\lambda^2_n
\frac{\nu t}{a^2} \big),
\end{equation}
and the mixed $r,t$ derivative yields
\begin{equation}\label{dtdrstart}
\frac{\partial^2 v_z(r,t)}{\partial t \partial r} = - \frac{2G}{\rho a}\sum_{n=1}^{\infty} \frac{J_1(\lambda_n
\frac{r}{a})}{J_1(\lambda_n)} \exp \big(-\lambda^2_n \frac{\nu
t}{a^2} \big).
\end{equation}
Here, $J_0, J_1$ denote the Bessel functions of order zero and one,
respectively, while $\lambda_n$ is the $n$th positive zero of $J_0$:
$J_0(\lambda_n) = 0$, and $J'_0(x) = -J_1(x)$. Note both the dynamic and kinematic viscosities appear 
in Eq. (\ref{start}).  We have used the fact that $\mu = \rho \nu$ in order to expose the dependence on the fluid density
in the derivatives in Eqs. (\ref{dtstart}, \ref{dtdrstart}).

\begin{figure}[htb]
\begin{center}
\includegraphics[width=0.70\textwidth]{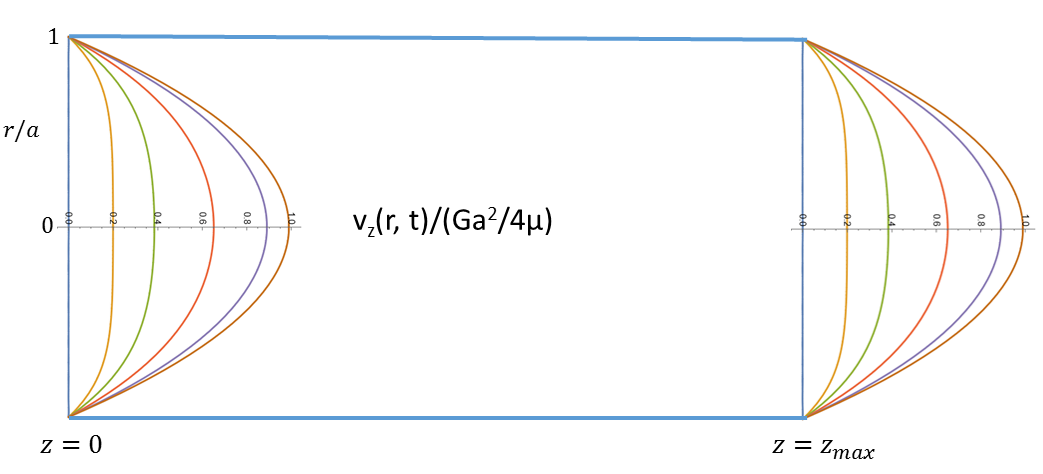}
\end{center}
\caption{\label{pipestartflow} Time-dependent velocity profile for
the starting flow problem in a cylindrical pipe of radius $a$,
within a region $0 \leq z \leq z_{max}$ of fully developed flow.
Fluid is initially at rest and the axial velocity $v_z$ Eq.
(\ref{start}) evolves to the final stationary parabolic profile of
Poiseuille flow for dimensionless times $\tau = \frac{\nu t}{a^2}
\simeq 1$. The instantaneous velocity is the same for all $z$ within
this region: here we display it at the entrance $z=0$ and exit
$z=z_{max}$ of the region of study. Sequence of velocity profiles
corresponds to $\tau = 0, 0.05, 0.1, 0.2, 0.4, 0.8$}
\end{figure}

It is convenient to plot the normalized axial velocity profile
$v_z(r,t)/(Ga^2/4\mu)$ using Eq. (\ref{start}), calculated across
the relative pipe diameter $0 \leq r/a \leq 1$ and for various
dimensionless times $\tau = \frac{\nu t}{a^2}$, see Fig
\ref{pipestartflow}. These curves are superposed on the entrance to,
and exit from, a linear section $0 \leq z \leq z_{max}$ of a
straight cylindrical pipe of radius $a$. Since we assume fully
developed flow within this region, the instantaneous velocity
profile is the same at both the entrance and exit. The no-slip
boundary condition implies the velocity is zero at the cylinder wall
$r=a$.

%

\subsection{\label{sec:startentropyprod} Starting flow in a pipe: the entropy production per unit volume}

\begin{figure}[htb]
\begin{center}
\includegraphics[width=0.65\textwidth]{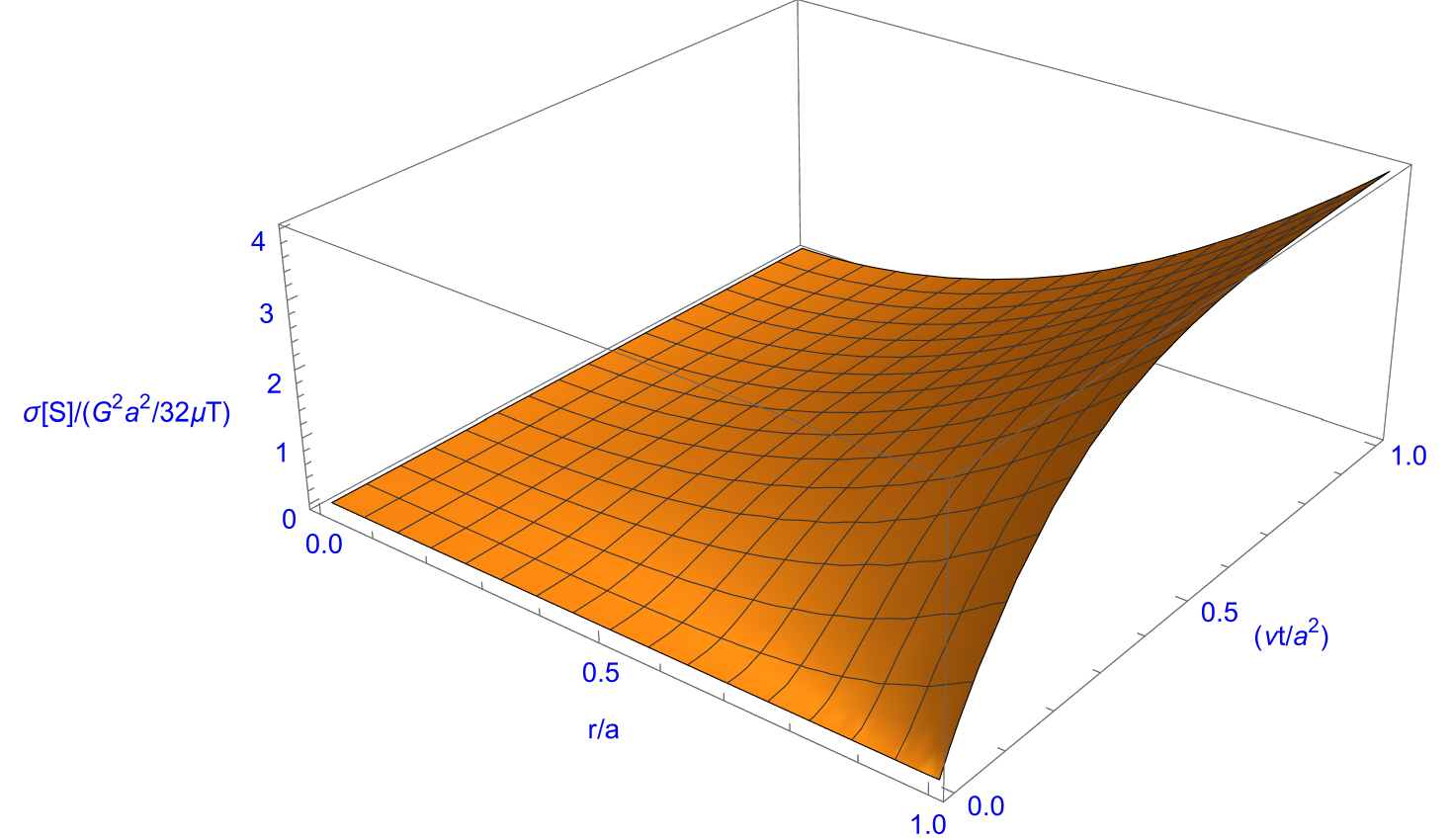}
\end{center}
\caption{\label{entropyprod} Plot of the viscous entropy production
per unit volume normalized as indicated: $\sigma^{visc}[S]/(G^2
a^2/32 \mu T)$ for a cylindrical pipe of radius $a$; see Eq.
(\ref{viscous}) and Eq. (\ref{drstart}). See text for discussion.}
\end{figure}

We now make the general expression for the bulk (volume) entropy
production Eqs. (\ref{source2c},\ref{visc}) explicit for the
starting flow problem. Thus, from Eqs.
(\ref{deviatoric},\ref{drstart}), this yields:
\begin{equation}\label{viscous}
\sigma^{visc}[S](r,t) = 2 \mu T^{-1} ||e_{zr}||^2 = \frac{\mu
T^{-1}}{2}  \Big(\frac{\partial v_z(r,t)}{\partial r}\Big)^2 \geq 0
\end{equation}
This function increases in time as the system relaxes to its
non-equilibrium stationary state, and the long time limit is given
by Eq. (\ref{PoisseuilleEntropy}).  A plot is given in Fig.
\ref{entropyprod} for a cylinder of radius $a$. The production is
zero on the axis of symmetry $r=0$ for all times $t \geq 0$, whereas
it increases in time from zero for all $r > 0$ and is a maximum at
the bounding wall $r=a$ (see Fig. \ref{entropyprod}). In the
starting flow problem, the entropy production per unit volume
initially is zero and \textit{increases} to a final positive value
for all $0 < r \leq a$ upon reaching the NESS.

%

\subsection{\label{sec:startingGEC} Starting flow in a pipe: the source term for $\Psi$}

\begin{figure}[htb]
\begin{center}
\includegraphics[width=0.65\textwidth]{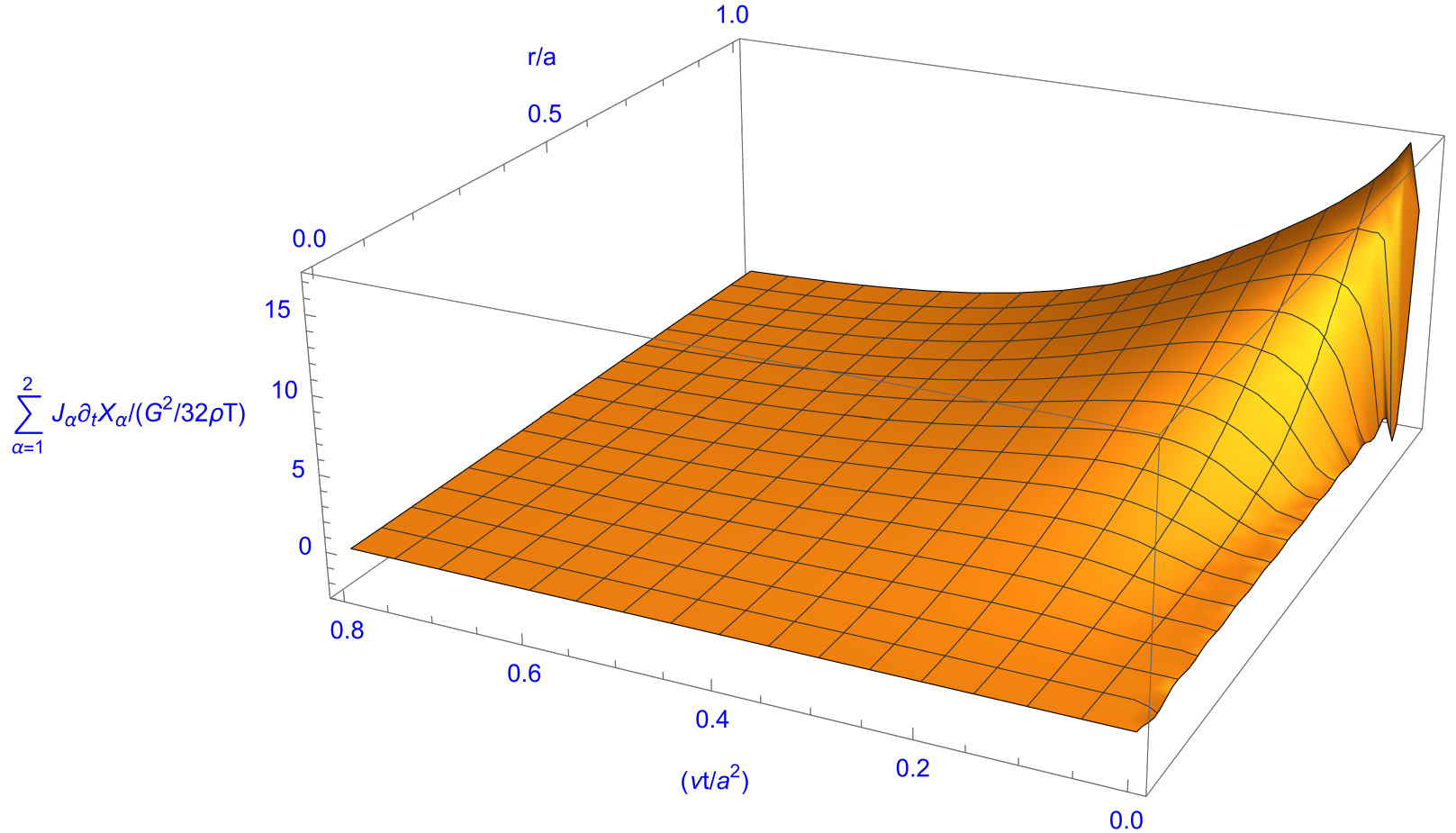}
\end{center}
\caption{\label{volGEC} Plot of normalized bulk volume contribution
$\sum_{\alpha} J_{\alpha}\,
\partial_t X_{\alpha}/(G^2/32 \rho T)$, see  Eq. (\ref{GECreduction}).
This is the \textit{time-derivative} of the viscous entropy
production Eq. (\ref{viscous}), see Fig. \ref{entropyprod}. }
\end{figure}

The bulk (volume) contribution to $\Psi$ is given above in
Eq.(\ref{GECreduction2}). As there is only the one velocity
component $v_z(r,t)$, that expression reduces and simplifies to give
\begin{eqnarray}\label{GECreduction}
\sum_{\alpha} J_{\alpha}\, \partial_t X_{\alpha} &=& +T^{-1}
d_{zr}\, \partial_t(v_{z,r}) = \frac{+T^{-1} \mu}{2}
\frac{\partial}{\partial t}\Big(\frac{\partial v_z(r,t)}{\partial r}
\Big)^2 = \frac{\partial}{\partial t}\, \sigma^{visc}[S](r,t) \geq
0.
\end{eqnarray}
This volume term is the time-derivative of the entropy production
(per unit volume)  Eq. (\ref{viscous}).  Since the latter is
positive and increases in time upon reaching the NESS for all $r>0$
(see Fig. \ref{entropyprod}), this quantity is also positive and
initially increases in time and then decreases monotonically in time
to zero as the system approaches its NESS, see Fig. \ref{volGEC}.
The initial increase in time to a maximum positive value is ever
more pronounced as one goes from the axis of symmetry $r=0$ to the
pipe wall $r=a$, see Fig. \ref{volGEC}. The integral of
Eq.(\ref{GECreduction2}) over the system volume must therefore also
be positive. In which case, we see that the GEC Eq.
(\ref{GECglobal}) does not hold.

\begin{figure}[htb]
\begin{center}
\includegraphics[width=0.70\textwidth]{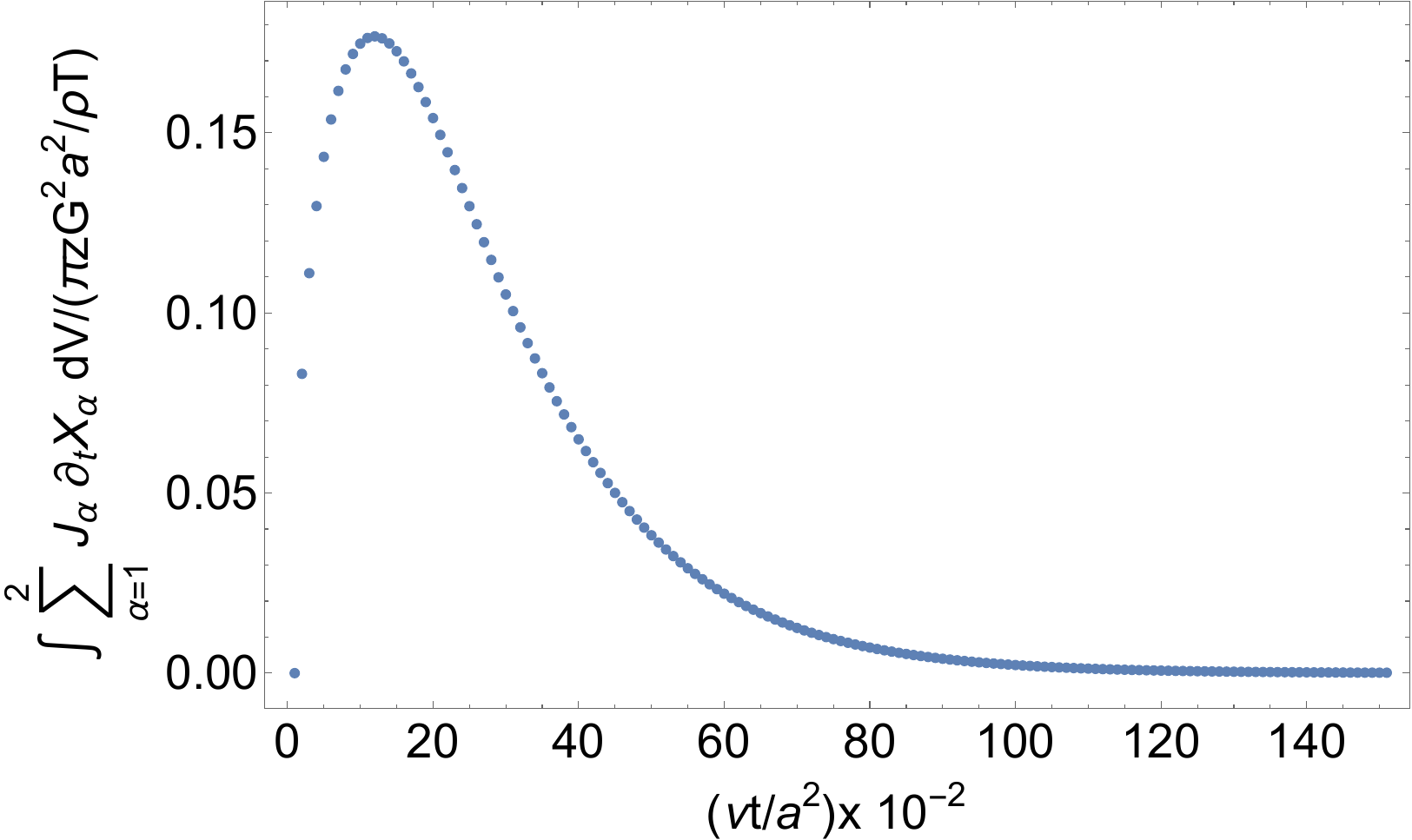}
\end{center}
\caption{\label{volumnAll} The volume integral of the complete bulk
volume contribution to the quantity $\Psi$: Eq. (\ref{JdX}) and
normalized as indicated; where $z$ is the length of pipe section
over which the flow is fully developed, $a$ is the pipe radius, $G$
the pressure gradient in the axial direction, $\rho$ is the fluid density and $T$ the
temperature.}
\end{figure}

\begin{figure}[htb]
\begin{center}
\includegraphics[width=0.70\textwidth]{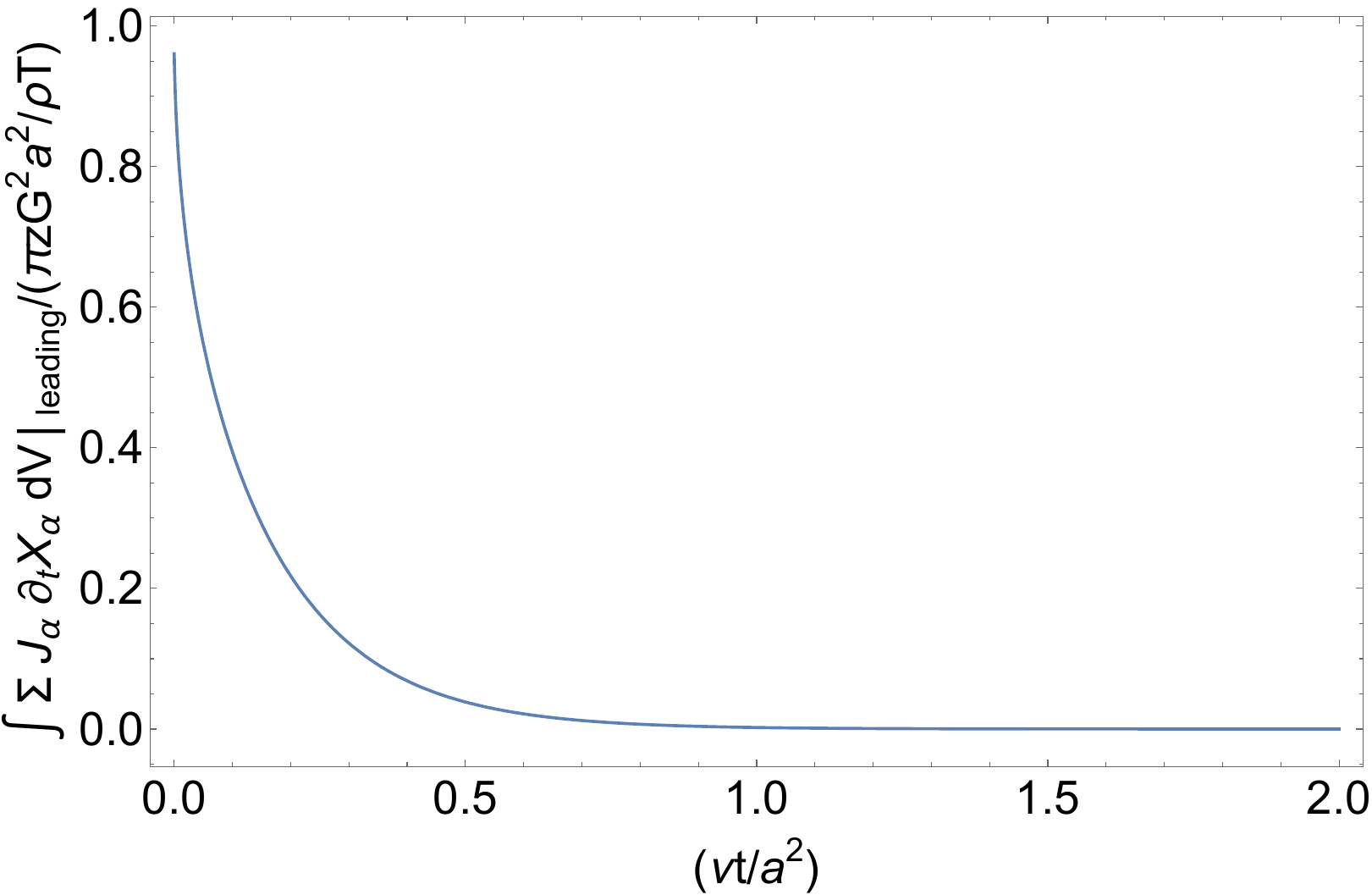}
\end{center}
\caption{\label{lead} The leading contribution to the volume
integral of the bulk volume contribution to $\Psi$: Eq.
(\ref{bulkleading}) and normalized as indicated. }
\end{figure}

We integrate this over the volume ($V_{cylinder} = \pi a^2 z_{max}$)
to obtain its contribution to the identity in Eq. (\ref{weakGEC}):
\begin{eqnarray}\label{JdX}
\int_V \sum_{\alpha} J_{\alpha}\,
\partial_t X_{\alpha}\,\, dV &=& 2\pi z_{max} \mu T^{-1} \int_0^a r
dr \, \frac{\partial v_z(r,t)}{\partial r} \, \frac{\partial^2
v_z(r,t)}{\partial t \partial r}\\
&=& 2\pi z_{max} \mu T^{-1} \int_0^a r dr \, \{ -\frac{G r}{2\mu} +
\frac{2 G a}{\mu}\sum_{n=1}^{\infty} \frac{J_1(\lambda_n
\frac{r}{a})}{\lambda_n^2 J_1(\lambda_n)}
\exp\big(-\frac{\lambda_n^2 \nu t}{a^2}\big)\}\nonumber \\
&\times&  \{ -\frac{2 G}{\rho a}\sum_{n=1}^{\infty}
\frac{J_1(\lambda_n \frac{r}{a})}{J_1(\lambda_n)}
\exp\big(-\frac{\lambda_n^2 \nu t}{a^2}\big)\} .
\end{eqnarray}
Evaluating this expression, we find it increases from zero to a
\textit{positive} and subsequently decays rapidly to zero from
above, see Fig. \ref{volumnAll}. Hence, the source term by itself in
Eq. (\ref{weakGEC}) behaves in a opposite and contrary sense to the
putative inequality. This strongly suggests that the surface
integral of the flow term must be sufficiently negative so as to
cancel and overcome this positive volume term, in order that the
inequality in Eq. (\ref{weakGEC}) be satisfied.
To this end, we first define and evaluate the \textit{leading}
contribution to Eq. (\ref{JdX}), which is defined by multiplying the
term $-\frac{G r}{2\mu}$ in the first set of braces by the complete
expression in the second braced factor. This yields
\begin{eqnarray}\label{bulkleading}
\int_V \sum_{\alpha} J_{\alpha}\,
\partial_t X_{\alpha}\,\, dV|_{leading} &=& \frac{2\pi
z_{max}T^{-1}G^2 a^2}{\rho} \sum_{n=1}^{\infty}
\frac{J_2(\lambda_n)}{\lambda_n J_1(\lambda_n)}
\exp\big(-\frac{\lambda_n^2 \nu t}{a^2}\big).
\end{eqnarray}
This is plotted in Fig. \ref{lead} for a cylinder of radius $a=1$
and shows that this quantity is positive and decreases monotonically
in time to zero. Note the normalization involving the model
parameters, and the dimensionless time units, see Eq.
(\ref{bulkleading}).

%

\subsection{\label{sec:startingboundary} Starting flow in a pipe: the boundary contributions}

We make explicit the boundary contributions to the evolution
criterion expression Eq. (\ref{weakGEC},\ref{surface}) for the
starting flow problem. Since the only nonvanishing component of the
velocity field is $v_z$, the expression Eq.
(\ref{weakGEC},\ref{surface}) can be written as as follows (in
the notation of \cite{GPbook}):
\begin{eqnarray}
\int_A \big(P_{ij} T^{-1} \partial_t v_i \big)n_j \, dA  &=&
\int_{A= caps} \big(p T^{-1} \partial_t v_z \big)n_z \,\, dA +
\int_{A=wall} \big(p_{zr} T^{-1}
\partial_t v_z \big)n_r \,\, dA,
\end{eqnarray}
where $n_z, n_r$ denote the (outward) axial and radial unit normals
to the boundary surface, respectively.  The first integral is over
the pair of cylinder entry and exit ``end-caps" located at $z=0$ and
at $z=z_{max}$, respectively, and the second integral is over the
cylinder wall located at $r=a$ (see Fig. \ref{pipestartflow}). We work these contributions out in more detail.

First deal with the contribution coming from the cylinder wall.
Note that the integral over the wall is at $r=a$:
\begin{equation}
\int_{A=wall} f(r,t) \,dA  = a \int_0^{2\pi} d\theta
\int_{z=0}^{z_{max}} dz f(a,t),
\end{equation}
and from inspection of Eq.(\ref{dtstart}), and since $J_0(\lambda_n)
= 0$, it is easy to see that the time-derivative of the velocity
vanishes at the wall: $\frac{\partial v_z(r,t)}{\partial t}|_{r=a} =
0$. So we can ignore the wall contribution. We will
have a finite contribution from the entry and exit end-caps. The
integrals over the two end caps are located at $z=0$ and at
$z=z_{max}$:
\begin{equation}
\int_{A=caps} dA = \int_{r=0}^{r=a} r\, dr\int_0^{2\pi} d\theta,
\end{equation}

\begin{figure}[htb]
\begin{center}
\includegraphics[width=0.70\textwidth]{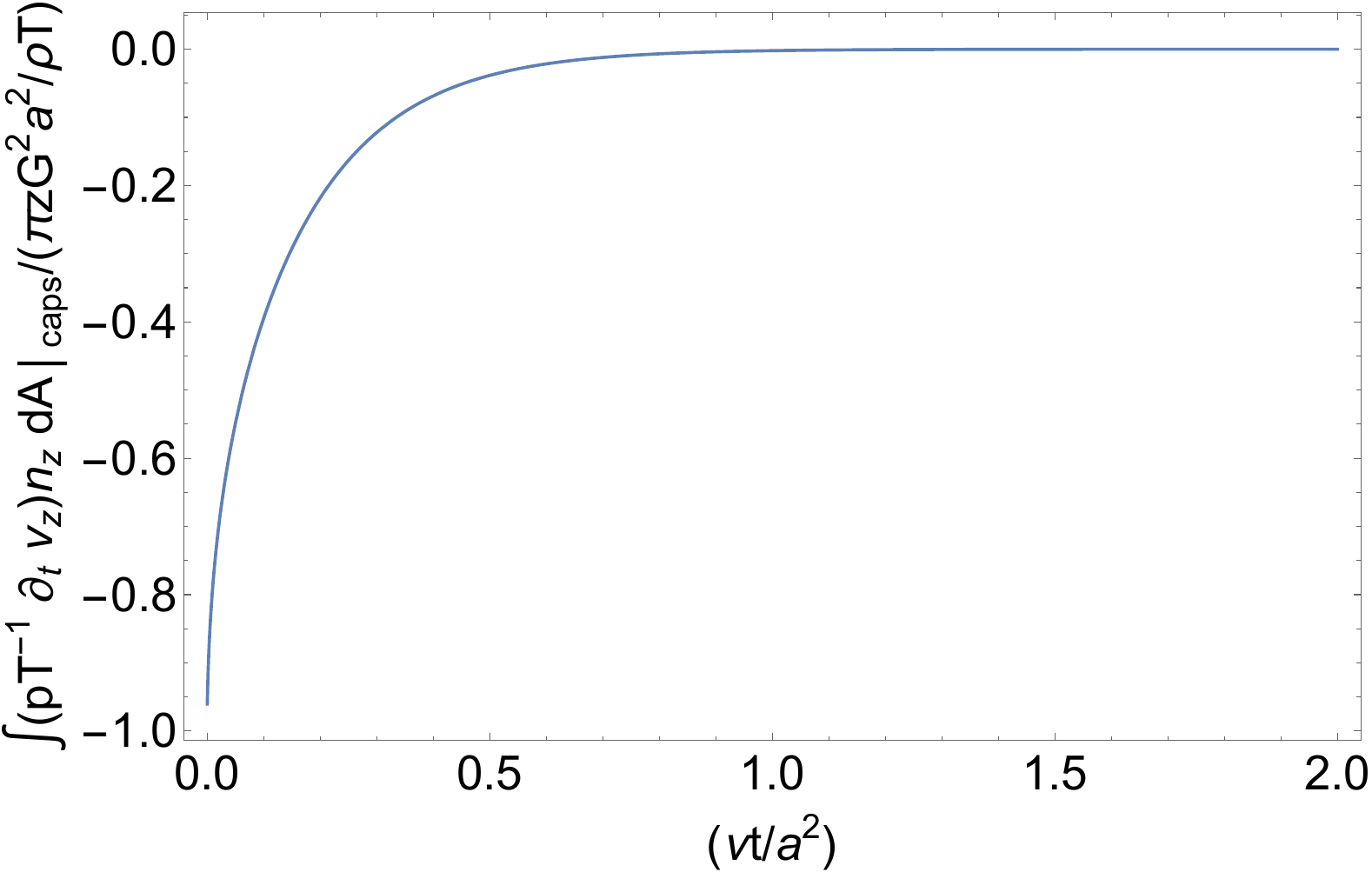}
\end{center}
\caption{\label{cylindercaps} Plot of Eq. (\ref{boundary5}),
normalized as indicated in the Figure.  This boundary contribution
comes exclusively from the cylinder entrance and exit surfaces (the
endcaps), is negative and increases to zero as the system approaches
its NESS. Compare to the curve in Fig. \ref{lead}.}
\end{figure}

We make the end-cap contribution explicit. First, note that
the external pressure applied to the pipe is given by
\begin{equation}
p(z) = p_0 -G z,
\end{equation}
where $p_0 = p(0)$ is the pressure applied at the input at $z=0$ and
decreases with distance from the input at constant rate $-G$, where
$G>0$ is constant. Note: this is how the pressure is specified in
the Poiseuille problem, see Batchelor. Then the contributions from
the two end-caps (located at $z=0$ and at $z=z_{max} = L$, where $L$ is the length of the pipe, see Fig.
\ref{pipestartflow}) are:

\begin{eqnarray}\label{boundary}
\int_{A= caps} \big(p T^{-1} \partial_t v_z \big)n_z \,\, dA  &=&
\int_{r=0}^{r=a} r\, dr\int_0^{2\pi} d\theta \, \Big( (p(z) T^{-1}
\partial_t v_z \big)n_z|_{z=0} + {(...)|_{z=z_{max}}} \Big),\\
&=& 2\pi T^{-1}\Big(p_0 n_z(0) + p(z_{max})n_z(z_{max})\Big)
\,\int_{r=0}^{r=a} r\, \frac{\partial v_z(r,t)}{\partial t} dr\\
&=& 2\pi T^{-1}\Big(-p_0  + (p_0 -Gz_{max})\Big) \,\int_{r=0}^{r=a}
r\, \frac{\partial v_z(r,t)}{\partial t} dr\\ \label{boundary4}
&=&
-2\pi T^{-1} Gz_{max} \,\int_{r=0}^{r=a} r\, \frac{\partial
v_z(r,t)}{\partial t} dr,\\ \label{boundary5} &=& -\frac{4 \pi
T^{-1} G^2 z_{max} a^2}{\rho}\sum_{n=1}^{\infty}
\frac{1}{\lambda^2_n} \exp \big(-\lambda^2_n \frac{\nu t}{a^2} \big)
\leq 0.
\end{eqnarray}
using the outward unit normals: $n_z(0) = -\hat{z}$, and
$n_z(z_{max}) = +\hat{z}$. Note the cylinder end-cap contribution to
the evolution criterion Eq. (\ref{weakGEC},\ref{surface}) is
\textsf{negative-definite}, see Fig. \ref{cylindercaps}.
In arriving at Eq. (\ref{boundary4}) we integrated the infinite
series term-by-term to get a closed form expression for the end-cap
contributions, using:
$$\int x^n J_{n-1}(x) dx = x^n J_n(x)$$
for $n=1$.

\subsection{\label{sec:startingGEC} Validation of the extended evolution criterion}

\begin{figure}[htb]
\begin{center}
\includegraphics[width=0.70\textwidth]{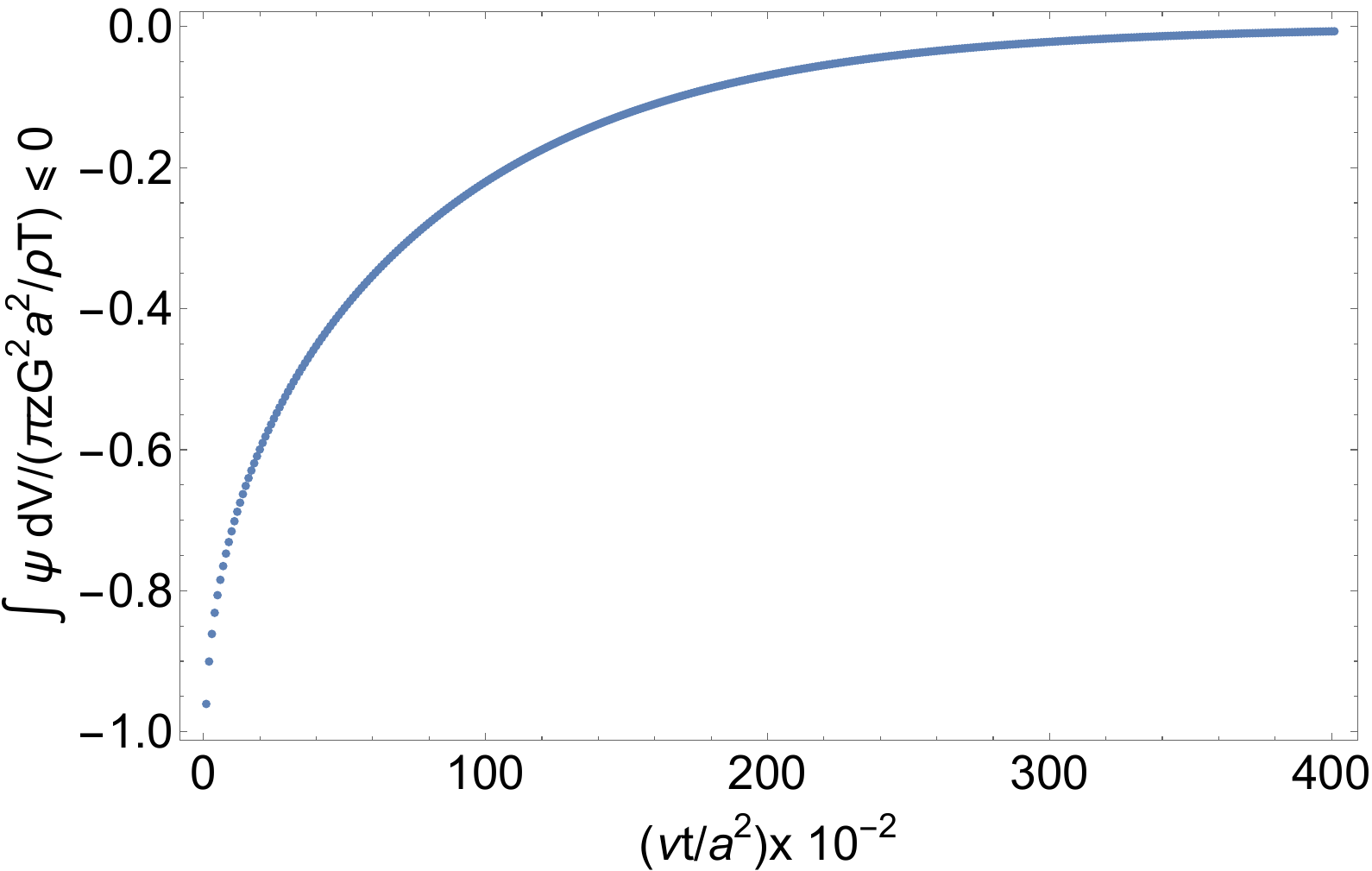}
\end{center}
\caption{Plot of $\int \Psi\, dV$ Eq. (\ref{weakGECF}) normalized as
indicated. This validates the \textit{extended} version of the
evolution criterion Eq. (\ref{weakGEC}). Both surface and volume
contributions are required in order that the inequality in Eq.
(\ref{weakGEC}) be obeyed. The GEC does not hold, but this more encompassing 
extended evolution criterion. }\label{ExtendedGEC}
\end{figure}

To demonstrate the validity of the extended GEC, we sum the leading term of the bulk
volume contribution and the boundary contribution using Eqs.
(\ref{bulkleading},\ref{boundary5}), and find that they mutually
cancel exactly:
\begin{equation}
\int_V \sum_{\alpha} J_{\alpha}\,
\partial_t X_{\alpha}\,\, dV|_{leading} + \int_{A= caps} \big(p T^{-1} \partial_t v_z \big)n_z
= 0,
\end{equation}
since
\begin{equation}
\frac{2}{\lambda_n} = \frac{J_2(\lambda_n)}{J_1(\lambda_n)}, \,
\forall \, n,
\end{equation}
and which follows from the functional relation $J_2(z) =
\frac{2}{z}J_1(z) - J_0(z)$ \cite{GradRyzhik}.  This exact
cancelation between the boundary (end-caps) and the leading bulk
volumn contributions leaves us finally with
\begin{equation}\label{weakGECF}
\int_V \Psi \, dV =  -8 \pi z_{max} T^{-1} \frac{G^2}{\rho}\int_0^a r dr \, \{\sum_{n=1}^{\infty} \frac{J_1(\lambda_n
\frac{r}{a})}{\lambda_n^2 J_1(\lambda_n)}
\exp\big(-\frac{\lambda_n^2 \nu t}{a^2}\big)\} \{\sum_{n=1}^{\infty}
\frac{J_1(\lambda_n \frac{r}{a})}{J_1(\lambda_n)}
\exp\big(-\frac{\lambda_n^2 \nu t}{a^2}\big) \} \leq 0.
\end{equation}
This is negative definite, and approaches zero from below as the
system relaxes to its NESS, see Fig. \ref{ExtendedGEC}.

The flow term (which appears in the boundary contribution) is
negative, whereas the source term (volume contribution) is positive,
but the former cancels the \textit{leading} positive contribution of
the latter, leaving behind a residual negative-definite volume term.
There is perfect cancelation of boundary (cylinder entry and exit
end-caps) and leading bulk at all matching temporal orders. We thus
confirm, in this model, that the inequality Eq. (\ref{weakGEC})
holds for the starting flow problem. But, in order that this be so,
both surface and volume contributions to the integral of $\Psi$ are
required. There is a partial cancelation between them that leaves a
negative definite remainder.

Note however, this is not the GEC as put forward by Glansdorff
and Prigogine. Indeed, the volume contribution is positive definite, its sign opposite to that predicted by the GEC, see Figure \ref{volumnAll}. 
The system evolution is sensitive to 
time-dependent boundary conditions, and the GEC
holds only for \textit{time-indepedent} boundary conditions
\cite{GP1964,GPbook}. 

\bigskip

\section{\label{sec:Numerical_approximation} Numerical approximation}
\subsection{Mathematical model.}

Let us consider an incompressible flow contained in a fix domain
$\Omega \in \mathbf{R}^{3}$ with boundary $\Gamma $ $\in
\mathbf{R}^{2}.$ Thus, in absence of body forces, the fluid motion is
governed by the Navier-Stokes equations in $\Omega $, which can be
written in conservative form as:

\begin{equation}
\rho \frac{\partial \mathbf{v}}{\partial t}+\rho \text{ }div(\mathbf{v}%
\otimes \mathbf{v})=div\mathbf{(-}p\mathbf{I+\tau )}\qquad for\text{ }%
\mathbf{x}\in \Omega \text{ }and\text{ }t>0  \label{eq1}
\end{equation}

\bigskip

coupled with the mass conservation equation, that in the
incompressible limit reduces to:

\begin{equation}
div\text{ }\mathbf{v}=0\qquad for\text{ }\mathbf{x}\in \Omega \text{ }and%
\text{ }t>0  \label{eq2}
\end{equation}

\bigskip

where $\mathbf{v}(\mathbf{x},t)$ and $p(\mathbf{x},t)$ are the
velocity and pressure fields; $\rho $ is the fluid density; and the
deviatoric component of the stress tensor, $\mathbf{\tau ,}$ is
defined as

\bigskip

\begin{equation}
\tau _{ij}(\mathbf{v})=\mu \left( \frac{\partial v_{i}}{\partial x_{j}}+%
\frac{\partial v_{j}}{\partial x_{i}}\right)   \label{eq3}
\end{equation}

\bigskip

where $\mu $ is the dynamic viscosity of the fluid.

\bigskip

The above system of equations has to be complemented by the initial
and boundary conditions in $\Omega $ and $\Gamma $, respectively.

\bigskip

\subsection{Numerical model.}

The numerical method used to solve the flow equations (\ref{eq1}-\ref{eq3}) was developed by Herreros and Ligüérzana \cite{Herreros2020} within the frame of the Finite Element Method and has been recently applied to the study of Dean vortices in curved pipes at low Reynolds numbers by Herreros and Hochberg \cite{HHPoF}. For the sake of completeness, it is summarized in the following, although the interested reader can find the details of its derivation and validation in \cite{Herreros2020} and \cite{HHPoF}.

\bigskip

In order to solve Navier-Stokes equations (\ref{eq1})-(\ref{eq3})
for a time step $\Delta t$, the fractional-step procedure proposed by Chorin \cite{Chorin} is followed. Concerning the spatial discretization, 3D linear tetrahedral have
been chosen because of the numerical efficiency and excellent
behaviour exhibited by these elements \cite{Peraire89,Zien1998}.
However, Babuska-Brezzi condition \cite{Babuska,Brezzi} does not
allow the use of the same order of interpolation for both velocity
and pressure unless a special stabilization technique is used
\cite{YangAtluri1984,ZienBookVol1,ZienBookVol3}. As was shown in
\cite{Chorin,Schneider,Ramaswamy1988,Wu,Codina1995,ZienCodina1995,Mabssout06,ZienBookVol2}, the proposed algorithm provides with the required stabilization.

\subsubsection{Step I: Fractional velocity discretization.}

It consists of solving explicitly:

\begin{equation}
\rho\frac{\partial }{\partial t}(v_{i})^{\ast n}+\rho\frac{\partial }{\partial
x_{j}}(v_{j}v_{i})^{n}-\frac{\partial }{\partial x_{j}}(\tau
_{ij})^{n}=0  \label{pasoI2}
\end{equation}

Note that equations (\ref{pasoI2}) contain all the advective terms.

The fractional linear momentum, equation (\ref{pasoI2}), is
discretized using the two-step Taylor-Galerkin algorithm
\cite{Mabssout06,Peraire86,PeraireZien86}, and the resulting system of equations is then solved using an iterative method of Jacobi type \cite{Hirsch}. See \cite{Herreros2020} and \cite{HHPoF} for the details on the numerical scheme.

The solution of (\ref{pasoI2}) will provide with a intermediate value for the velocity: $\rho \mathbf{v}^{\ast n}=\rho \mathbf{v}%
^{n}+\Delta \rho \mathbf{v}^{\ast n}.$

\subsubsection{Step II: Continuity equation discretization.}
\bigskip

Applying the standard Galerkin discretization, we then solve implicitly the following equation \cite{Herreros2020} \cite{HHPoF}: 

\begin{equation}
\frac{\partial }{\partial t}(\rho v_{i})^{\ast \ast n}+\frac{\partial }{%
\partial x_{j}}(\delta _{ij}p)^{n+1}=0  \label{pasoII2}
\end{equation}

which accounts for the pressure term, along with the incompressibility condition: 

\begin{equation}
div\text{ }\mathbf{v}^{n+1}=0  \label{continuity}
\end{equation}%

\bigskip

The resulting system of equations is solved using the conjugate
gradient method \cite{ZienBookVol1, Hirsch}, providing with $\Delta \rho \mathbf{v}^{\ast \ast n}=\rho \mathbf{v}^{n+1}-\rho 
\mathbf{v}^{\ast n}$.

\subsubsection{Step III: Velocity correction.}

\bigskip Once pressure has been computed in the previous step, the velocity increment, $\Delta \mathbf{v}^{\ast }$, must be corrected by
discretizing in space equation (\ref{pasoII2}) and solving the resulting system of equations by means of a Jacobi iterative
scheme \cite{Hirsch}.

As a result, $\Delta \mathbf{v}^{\ast \ast n}$ is obtained, which is added to
$\Delta \mathbf{v}^{\ast n}$, resulting in the total velocity
increment $\Delta \mathbf{v}^{n}$ \cite{Herreros2020} \cite{HHPoF}:

\begin{equation}
\Delta \mathbf{v}^{n}=\Delta \mathbf{v}^{\ast n}+\Delta
\mathbf{v}^{\ast \ast n}
\end{equation}%
and thus, the velocity at the following time step is obtained:

\begin{equation}
\mathbf{v}^{n+1}=\mathbf{v}^{n}+\Delta \mathbf{v}^{n}
\end{equation}

\subsubsection{Stability of the scheme.}

The maximum allowed time step in the solution of Navier-Stokes
equations is \cite{ZienBookVol3}:

\bigskip
\begin{equation}
C\leq \beta \left( \sqrt{\frac{1}{P_{e}^{2}}+\alpha }-\frac{1}{P_{e}}\right) 
\label{courant}
\end{equation}

where:

\qquad $C$ is the Courant number: $C=\frac{\Delta
t|\mathbf{v}|}{h_{e}}$

\qquad $P_{e}$ is the Peclet number: $Pe=\frac{\rho
h_{e}\mathbf{|v|}}{2\mu } $

\qquad $h_{e}$ is the element size (minimum height of each element
has been considered).

\qquad $\alpha =1$ when using the lumped mass matrix and $\alpha
=1/3$ when using the consistent mass matrix in the solution of steps
I and III.

\qquad $\beta $ is a safety coefficient, typically 0.85-0.9

The global time step limit is calculated as the minimum time step
allowed in the mesh. Equation (\ref{courant}) incorporates the
effects of viscosity via
Peclet number and, in order to account for non-linearities, a safety factor $%
\beta $ is considered.

\subsection{\label{sec:NumericalC} Numerical computation of the extended evolution criterion}

In order to compute equation (\ref{weakGEC}) using the basic theory of Finite Element Method, the surface integral (\ref{surface}) in matrix form (see Appendix \ref{sec:matrix} for the details) is calculated at $t^{n+1}$ as: 

\begin{eqnarray}\label{surfaceFEM}
\bigg[\int_A \bigg(p T^{-1}\, (\partial_t \bm{v}) \, \bm{n}^T -T^{-1} (\partial_t \bm{v}) \, \mathbf{D} \,
\bm{n}^T \bigg) dA \bigg]^{n+1} &=& 
T^{-1} \bigg[\sum_{e_{b}} \bigg({p_{e}}^{n+1} \frac{\Delta\mathbf{{v}_{e}}^{n}}{\Delta t} \mathbf{n_{e}}^{T}-\frac{\Delta\mathbf{{v}_{e}}^{n}}{\Delta t} \mathbf{{D_{e}^{n+1}} }\mathbf{n_{e}}^{T}\bigg)\Gamma_{e_{b}}\Bigg] 
\end{eqnarray}

where
\begin{equation}
\Delta \mathbf{v_{e}}^{n}=\mathbf{v_{e}}^{n+1}-\mathbf{v_{e}}^{n}
\end{equation}

\begin{equation}
\mathbf{{D_{e}^{n+1}}} = \mu \big(\nabla \mathbf{N}^{T}\mathbf{v^{n+1}} + {\nabla}^T \mathbf{N} \big(\mathbf{{v^{n+1}}}\big)^{T} \big)
\end{equation}

The summation in Equation (\ref{surfaceFEM}) is extended to the boundary elements, assuming $T$ constant and where $\mathbf{D_{e}}$ is the deviatoric stress tensor at the element level, $\mathbf{N}^{T}$ is the FEM interpolating shape function, $\Gamma_{e_{b}}$ is the surface of the tetrahedra's facets at the boundaries and the subindex ${e_b}$ stands for \textit{boundary element}. 

The discretized form of the bulk volume contribution (\ref{GECreduction0}) is computed at $t^{n+1}$ as:
\begin{eqnarray}\
\bigg[\int_V \sum_{\alpha} J_{\alpha}\, \partial_t X_{\alpha} \, dV\bigg]^{n+1}  =\bigg[ \int_V \bigg(T^{-1} \, t_r [
\mathbf{D} \, (\partial_t \bm{\nabla} \bm{v})] + \rho T^{-1} \bm{v} \, \bm{\nabla}\bm{v} \, (\partial_t \bm{v})^T\Bigg)\, dV \bigg]^{n+1}= \\
\label{volumeFEM} = T^{-1} \bigg[\sum_{e} \bigg(t_r [\mathbf{{D_{e}^{n+1}}} \frac{\nabla \mathbf{N}^{T}\mathbf{v^{n+1}}-\nabla \mathbf{N}^{T}\mathbf{v^{n}}}{\Delta t}] + \rho \,\mathbf{v_{e}}^{n+1} \nabla \mathbf{N}^{T}\mathbf{v^{n+1}} \left[\frac{\Delta\mathbf{{v}_{e}}^{n}}{\Delta t}\right]^T\bigg)\Omega_e \Bigg]  
\end{eqnarray}

The summation in Equation (\ref{volumeFEM}) are extended to the tetrahedra's volumes, $\Omega_e$, where the subindex $e$ stands for \textit{element}.

Therefore, the numerical solution of Eq. (\ref{weakGEC}) for $t^{n+1}$ can be expressed as: 

\begin{eqnarray}\label{EGEC-FEM}
\bigg[\int_V \Psi \, dV \bigg]^{n+1}= T^{-1} \bigg[\sum_{e_{b}} \bigg({p_{e}}^{n+1} \frac{\Delta\mathbf{{v}_{e}}^{n}}{\Delta t} \mathbf{n_{e}}^{T}  - \frac{\Delta\mathbf{{v}_{e}}^{n}}{\Delta t} \mathbf{{D_{e}^{n+1}} }\mathbf{n_{e}}^{T}\bigg)\Gamma_{e_{b}}  + \\  +  \sum_{e} \bigg(t_r [\mathbf{{D_{e}^{n+1}}} \frac{\nabla \mathbf{N}^{T}\mathbf{v^{n+1}}-\nabla \mathbf{N}^{T}\mathbf{v^{n}}}{\Delta t}] + \rho \,\mathbf{v_{e}}^{n+1} \nabla \mathbf{N}^{T}\mathbf{v^{n+1}} \left[\frac{\Delta\mathbf{{v}_{e}}^{n}}{\Delta t}\right]^T\bigg)\Omega_e  \Bigg]
\end{eqnarray}

It is important to note here the impossibility to compute above equation for the initial time, since for each time, $t^{n+1}$, the values of the velocities in the previous time, $t^{n}$, must be known, condition which is obviously not fulfilled for $t=0$. Therefore, the value of $\int_V \Psi \, dV$ is set null for the initial time step. This fact will be useful to interpret some of the results presented in the following sections, since the first computed result will correspond to $t=\Delta t$, where $\Delta t$ is limited by the stability condition (\ref{courant}). 
\section{\label{sec:cylinder} Extended evolution Criterion for the Poiseuille flow}

In order to asses the performance of the numerical model in computing the proposed extended evolution criterion (\ref{weakGEC}), the starting flow problem, whose analytical solution was presented in Section \ref{sec:startPoiseuille}, is numerically solved.

\begin{figure}[!htb]
\begin{center}
\includegraphics[width=0.70\textwidth]{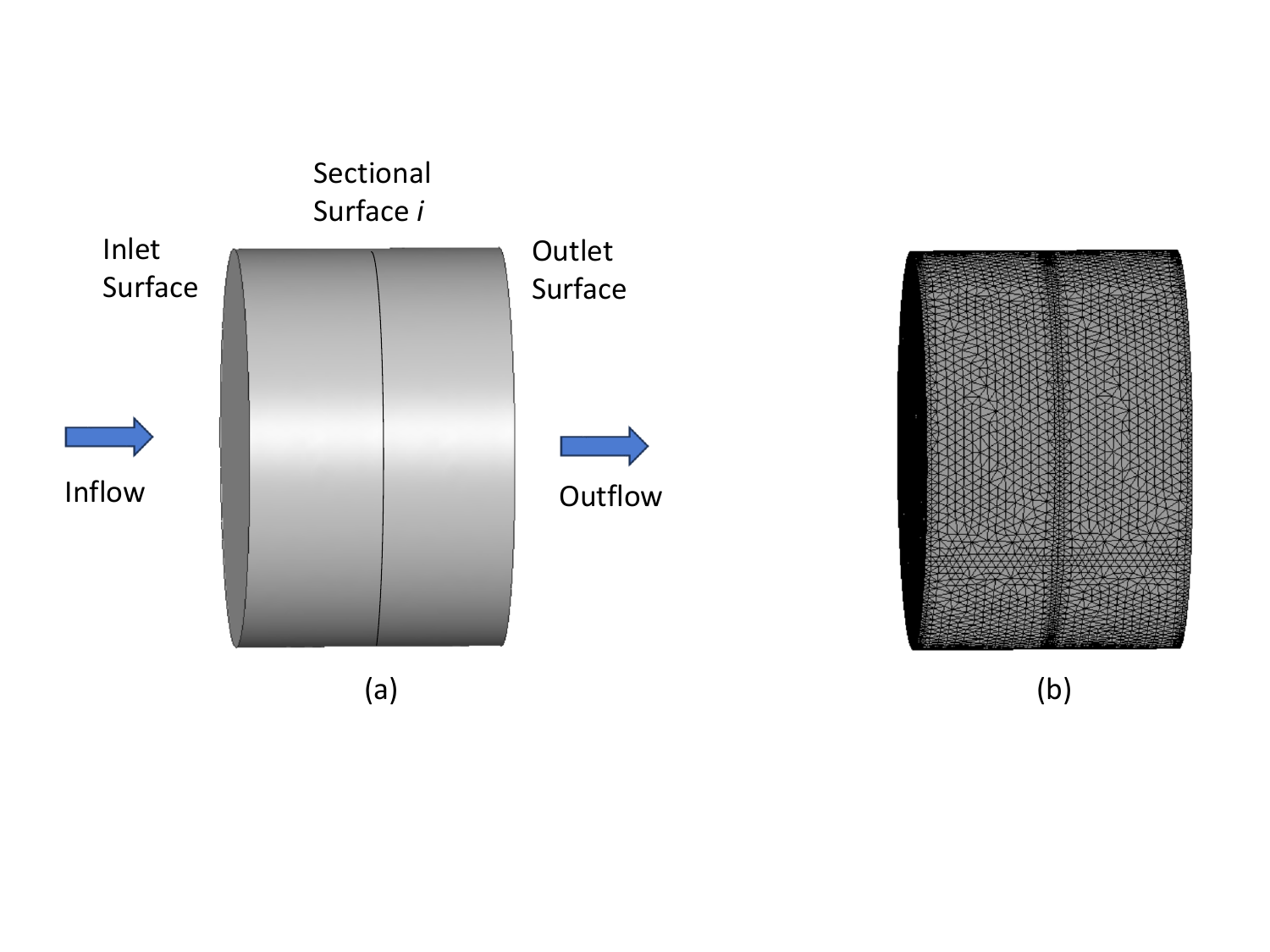}
\end{center}
\caption{Poiseuille flow: (a) Geometry: inlet, outlet and control surface \textit{i}; (b) Computational mesh of 85093 linear tetrahedra and 18670 nodes.}\label{Figure9}
\end{figure}

The layout of the problem is summarized in Figure \ref{Figure9}a. A short cylinder of radius $a = 1$ m and length $z_{max} = 1.33$ m is considered. 

\begin{figure}[!htb]
\begin{center}
\includegraphics[width=0.70\textwidth]{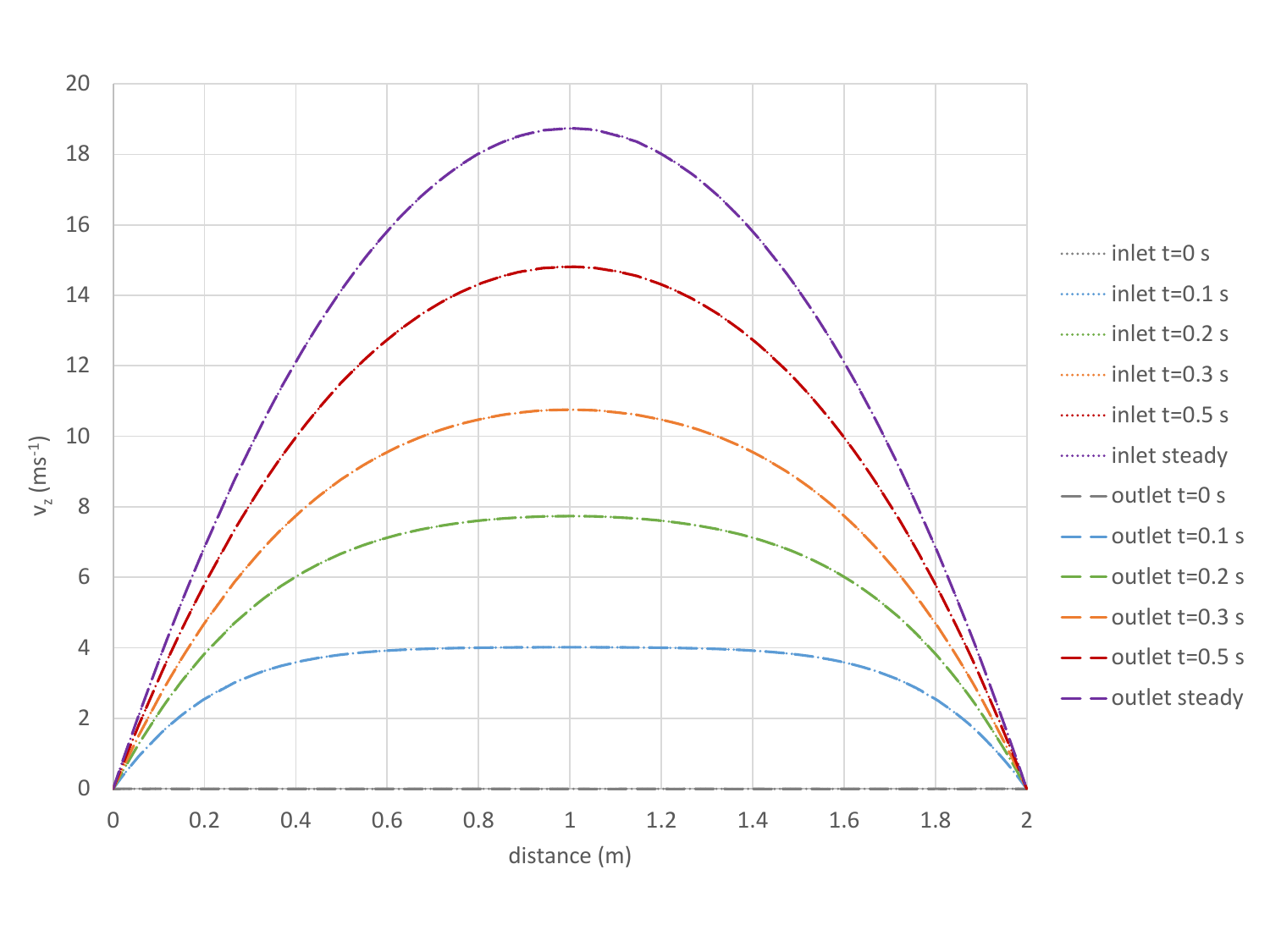}
\end{center}
\caption{Poiseuille flow: boundary conditions for the axial velocity at inlet and outlet where the curve color indicates the time and the curve marker indicates the surface. The analytical expression of the curves is given by Eq. (\ref{start}). Distance is measured over the sectional diameter.}\label{Figure10}
\end{figure}

A 3D computational mesh of 85093 linear tetrahedra and 18670 nodes is used for the computation (see Figure \ref{Figure9}b). The time-dependent boundary condition (\ref{start}) for the axial component of the velocity, ${v_{z}}$, is imposed at inlet and outlet surfaces, as shown in Figure \ref{Figure10}, while the secondary velocity, normal to the axial direction, is set null, ${v_{x}=v_{y}=0}$. Pressure $p=0$ is prescribed at the outlet, and the non-slip boundary condition, $\mathbf{v}=0$, is imposed at the pipe's walls. Initial conditions $\mathbf{v}=0$ and $p=0$, for velocity and pressure respectively, are considered.

\begin{figure}[!htb]
\begin{center}
\includegraphics[width=0.70\textwidth]{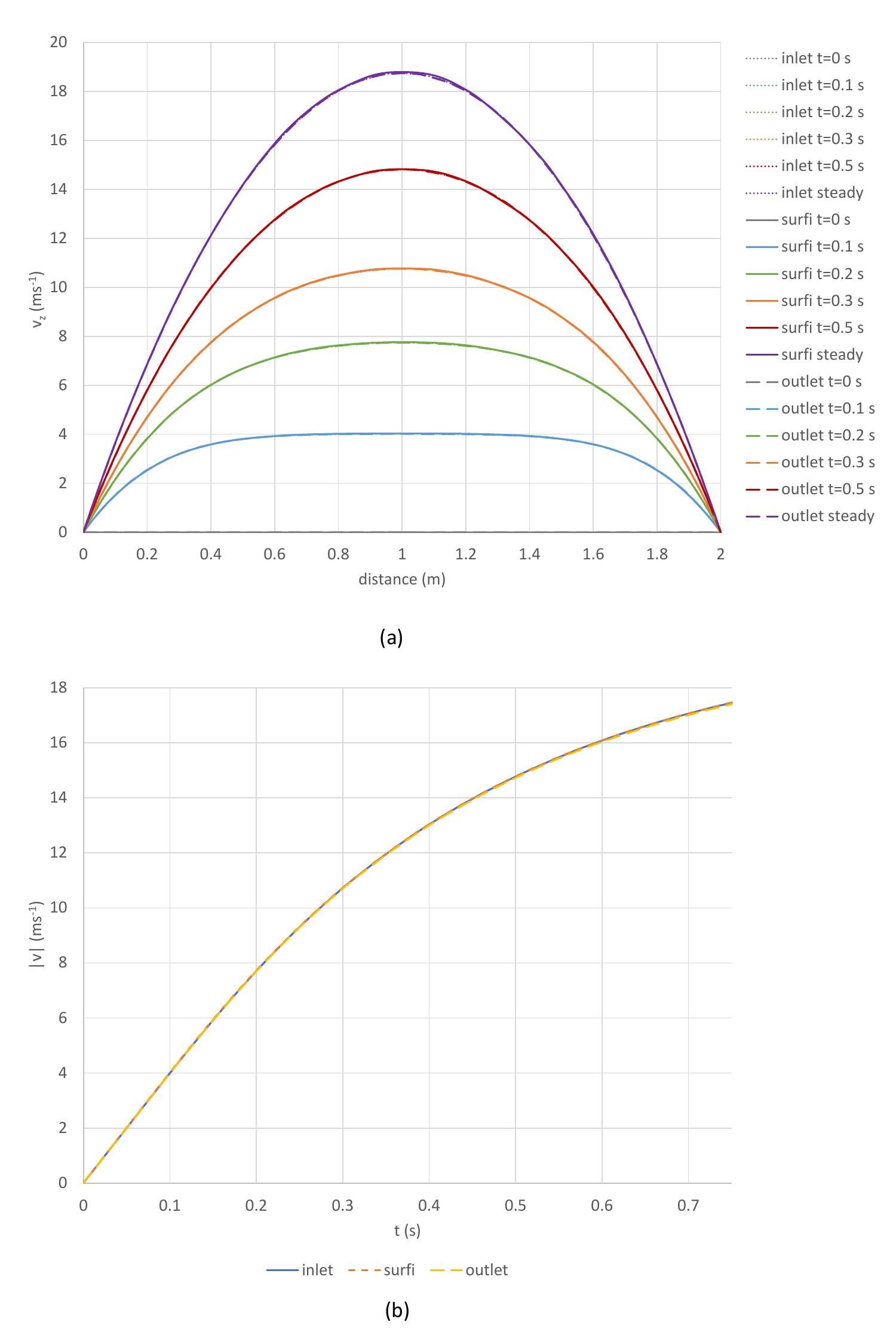}
\end{center}
\caption{Poiseuille flow: (a) axial component of velocity at inlet, outlet and control surface \textit{i}; same curve color indicates same time while same curve marker indicates same surface; distance is measured over the sectional diameter; (b) evolution of the modulus of velocity at the pipe's central axis for inlet, outlet and control surface \textit{i}.}.\label{Figure11}
\end{figure}

\begin{figure}[!htb]
\begin{center}
\includegraphics[width=0.70\textwidth]{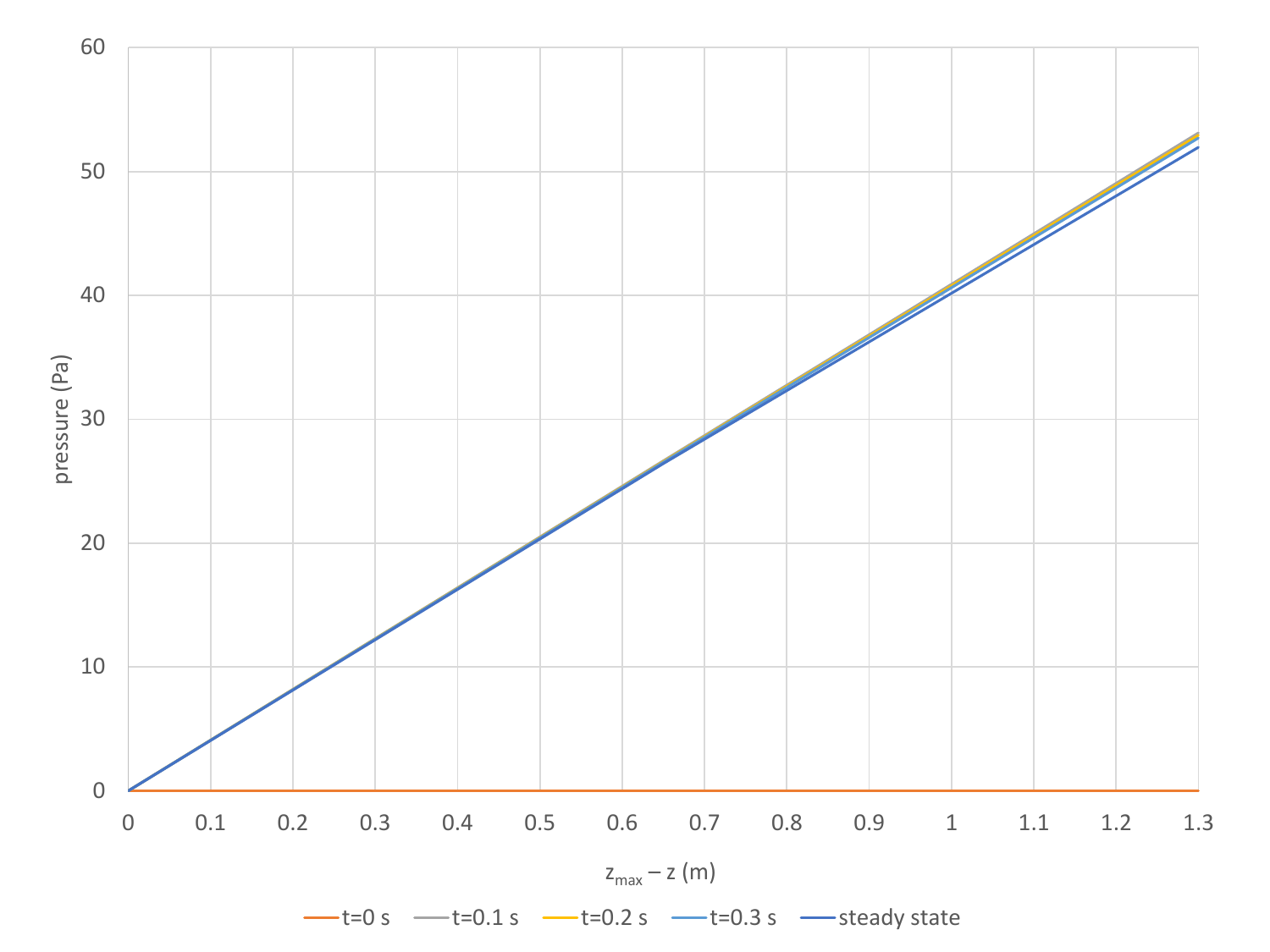}
\end{center}
\caption{Poiseuille flow: pressure variation along the pipe's central axis}\label{Figure12}
\end{figure}

The fluid is considered to be newtonian and incompressible, with density $\rho = 1$ kg m$^{-3}$ and dynamic viscosity $\mu = 0.5$ Pa s. The time step used in the computation is $\Delta t = 0.5 \cdot 10^{-4}$ s. The results obtained for the evolution in time of the velocity inside the pipe for the three sections considered are shown in Figure \ref{Figure11}. The pressure evolution along the pipe's central axis is depicted in Figure \ref{Figure12}.

\begin{figure}[!htb]
\begin{center}
\includegraphics[width=0.70\textwidth]{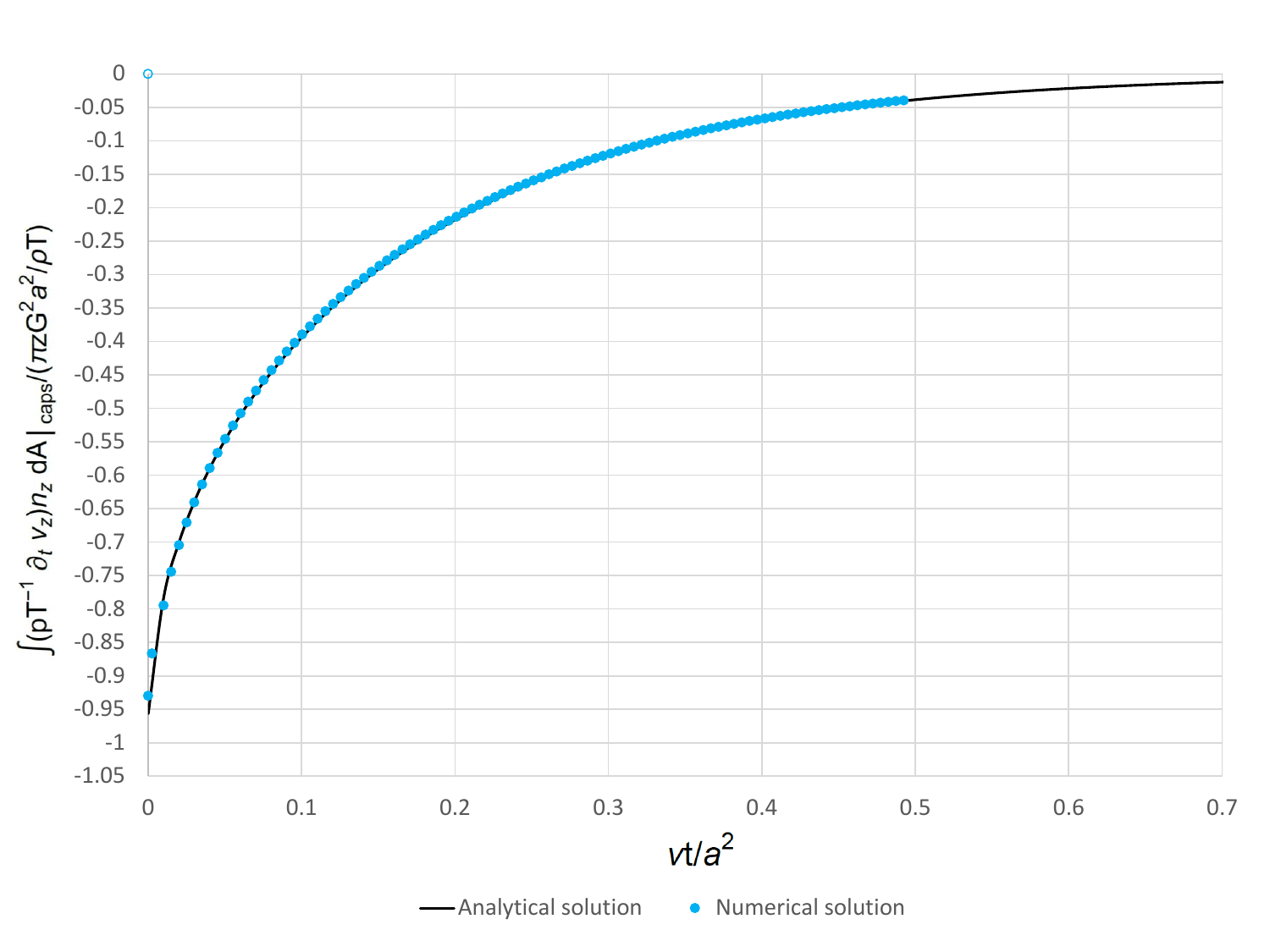}
\end{center}
\caption{Negative surface contribution: Normalized solution of Eq. (\ref{surfaceFEM}): analytical vs numerical}\label{Figure13}
\end{figure}

\begin{figure}[!htb]
\begin{center}
\includegraphics[width=0.70\textwidth]{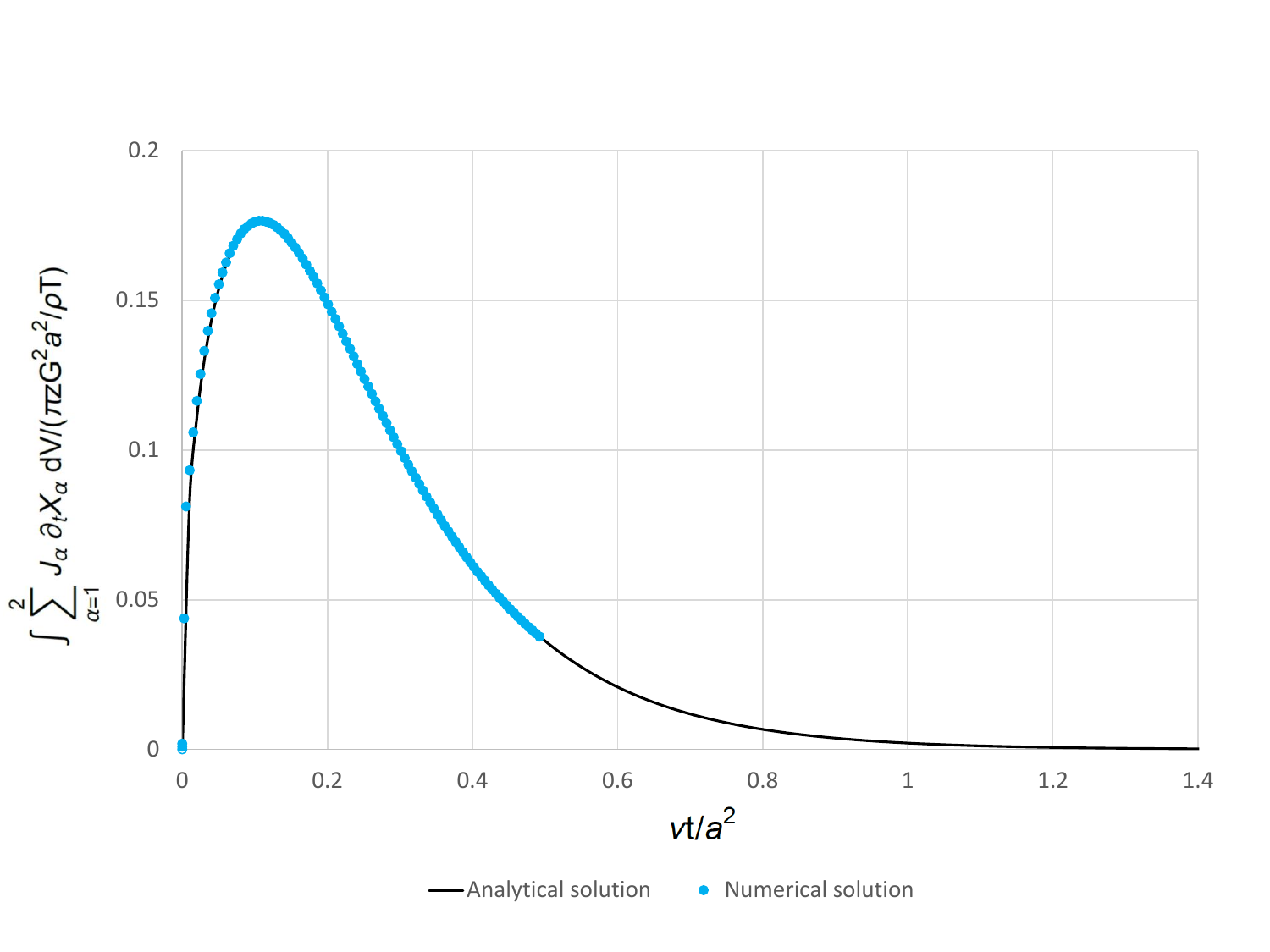}
\end{center}
\caption{Positive volume contribution: Normalized solution of Eq. (\ref{volumeFEM}): analytical vs numerical}\label{Figure14}
\end{figure}

\begin{figure}[!htb]
\begin{center}
\includegraphics[width=0.70\textwidth]{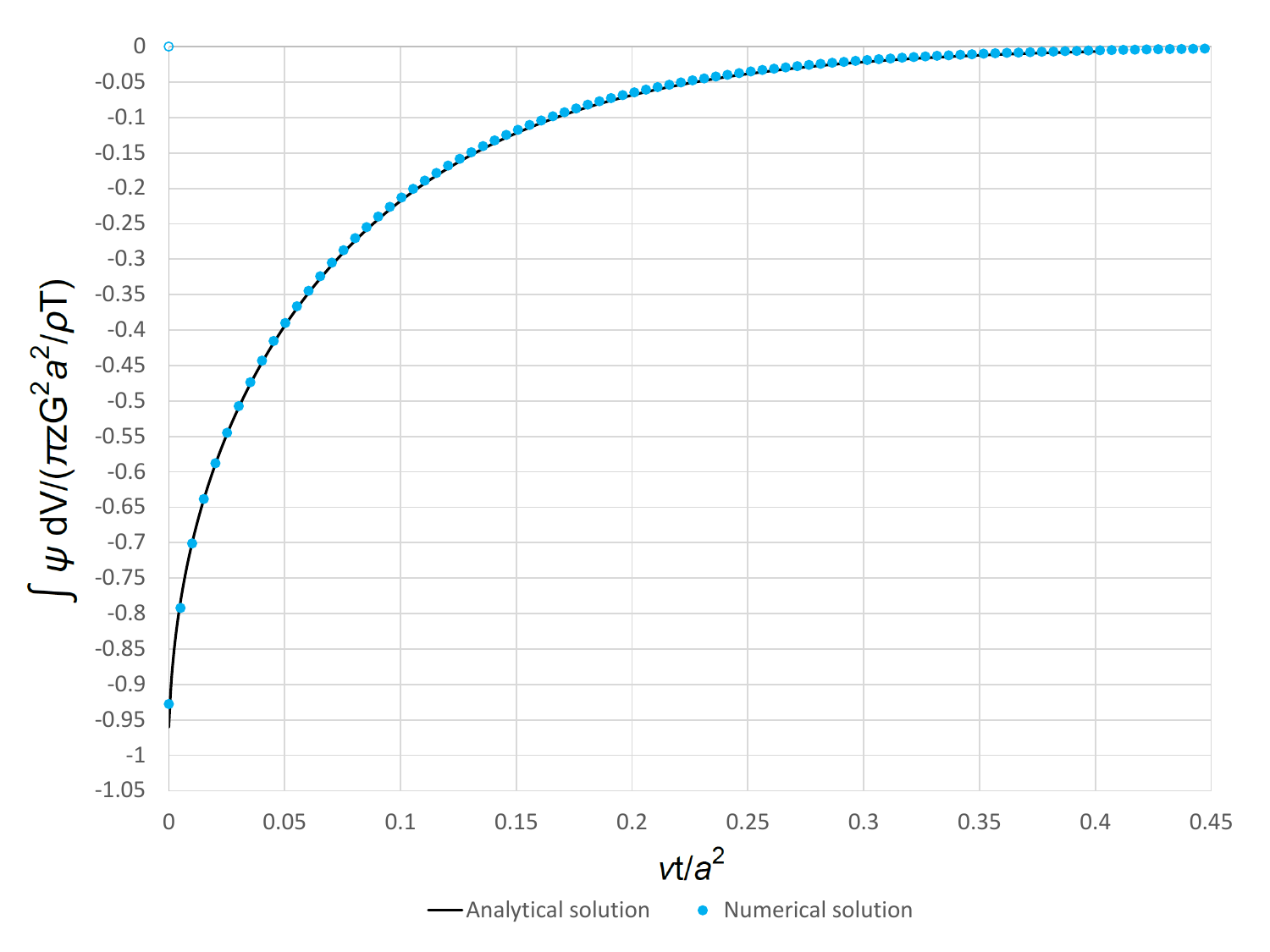}
\end{center}
\caption{Full negative contribution: Normalized solution of Eq. (\ref{EGEC-FEM}): analytical vs numerical }\label{Figure15}
\end{figure}

The analytical solution presented in Section \ref{sec:startPoiseuille} for the starting flow is derived under the assumption of a fully developed flow. Therefore, this condition must be fulfilled in the computational model as well. Figure \ref{Figure11}a shows the velocity profile for the axial velocity in the control section \textit{i}, located at a distance $z_{max}/2$ from the inlet, in comparison with the boundary conditions at inlet and outlet. Figure \ref{Figure11}b depicts the evolution of the velocity modulus in time at the center on the three considered sections. As shown in \ref{Figure11}, the solution for the velocity inside the volume, at section \textit{i} is identical to the imposed boundary conditions at the inlet and outlet surfaces, proving that the flow inside the pipe is fully developed. 
Once the values of velocity and pressure inside the domain, for each time step, are determined, the evaluation of the extended evolution criterion can be carried out. Figures \ref{Figure13}, \ref{Figure14} and \ref{Figure15} show the normalized curves obtained from computing Eqs.(\ref{surfaceFEM}), (\ref{volumeFEM}) and (\ref{EGEC-FEM}) respectively. It can be observed the perfect accordance between numerical and analytical solutions for the separate (surface and volume) and full contributions. Only a slight discrepancy is observed in Figs. \ref{Figure13} and \ref{Figure15} at $t=0$, due to the impossibility of numerically computing Eq. (\ref{EGEC-FEM}) for $t=0$.

Therefore, we confirm numerically what was previously demonstrated analytically in Section \ref{sec:startPoiseuille}: the inequality in Eq. (\ref{weakGEC}) \textit{does} holds for the starting flow problem. But, as discussed before, this is not the GEC as firstly stated by Glansdorff and Prigogine, since the evolution of this system involves the time-dependent boundary conditions, and the GEC, in its original formulation, holds only for \textit{fixed} boundary conditions \cite{GP1964,GPbook}. Thus, the extended evolution criterion given in (\ref{weakGEC}) is required to properly account for the time-dependent boundary contributions.

%

\section{\label{sec:helical} Evolution Criterion for Helical Pipe Flow}

In former sections we have demonstrated that the GEC, originally formulation by Glansdorff and Prigogine for systems under fixed boundary conditions, does not hold for the particular case of a fully developed flow of Poiseuille type with time-dependent boundary conditions. However, to generalize this statement we need to test the extended evolution criterion on a more general case: a duct with non-zero curvature and pitch (i.e. a helical pipe), where the flow is not fully developed,  subjected to time-dependent conditions at the boundaries. To do that, we will fist analyze the case of a curved pipe, with curvature and pitch and fixed boundary conditions and will compare with the results obtained for the same pipe subjected to time-dependent boundary conditions.

\subsection{\label{sec:helical-fixed} Helical flow under fixed boundary conditions}

\begin{figure}[!htb]
\begin{center}
\includegraphics[width=0.70\textwidth]{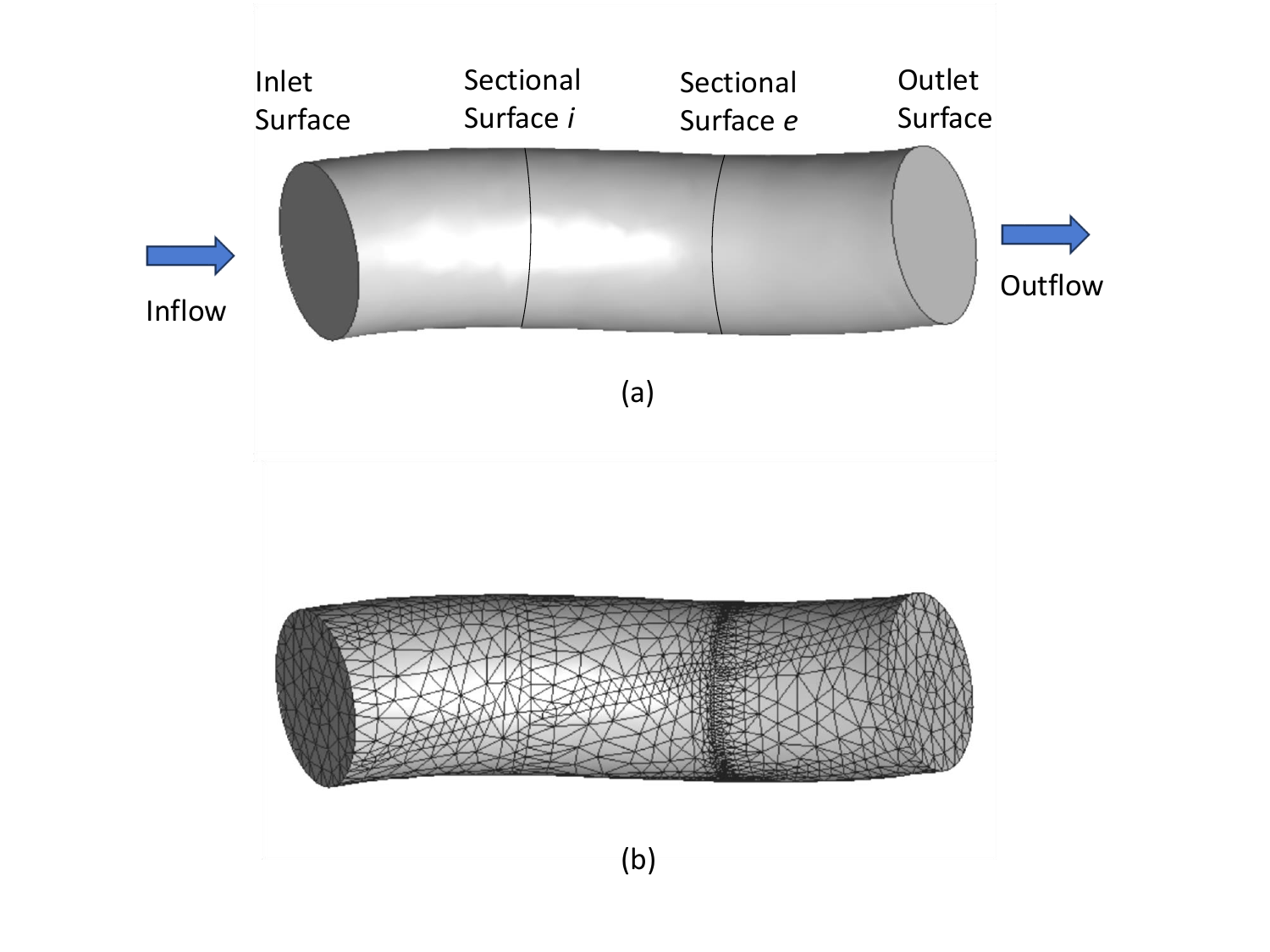}
\end{center}
\caption{Helical flow: (a) Geometry: inlet, outlet and inner control surfaces \textit{i} and \textit{e}; (b) Computational mesh of 48314 linear tetrahedra and 9175 nodes.}\label{Figure16}
\end{figure}

The setup of the problem is schematized in Figure \ref{Figure16}a. A helical pipe of sectional radius $a = 1$ m, dimensionless curvature $\kappa =a/R = 0.5$ and helical pitch $pitch = h/R = 10$, is considered for the analysis, where $h$ is the helical pitch (20 m) and $R$ is the radius of curvature (2 m). The length of the pipe is $3h/10$ and two internal surfaces are considered as control sections: surface \textit{i} located at a distance $h/10$ from the inlet and surface \textit{e}, at $2h/10$ from the inlet.

\begin{figure}[!htb]
\begin{center}
\includegraphics[width=0.70\textwidth]{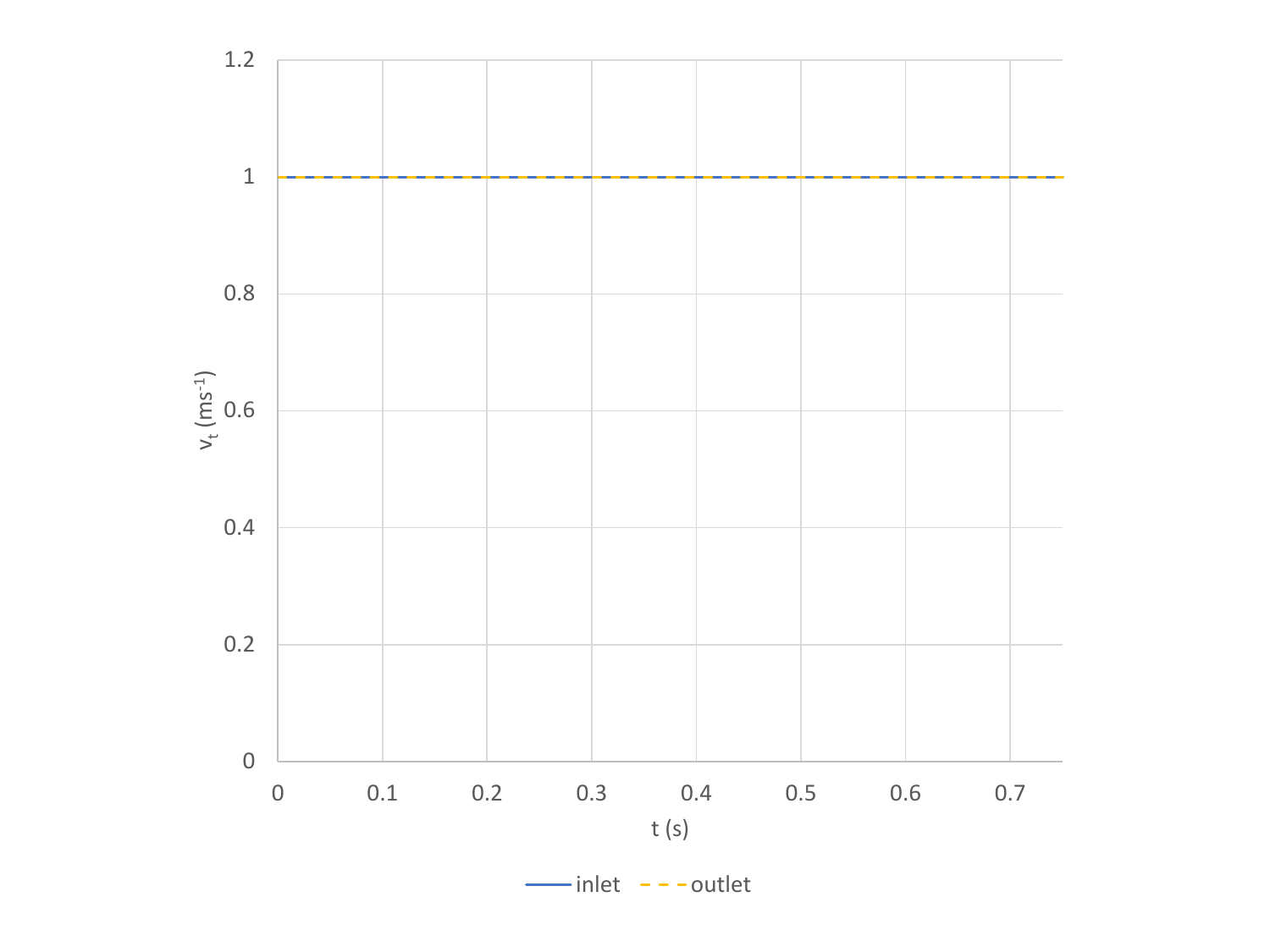}
\end{center}
\caption{Helical flow: Fixed boundary conditions for the axial velocity at inlet and outlet.}\label{Figure17}
\end{figure}

A 3D computational mesh of 48314 linear tetrahedra and 9175 nodes is used for the computation (see Figure \ref{Figure16}b). The boundary condition for the axial component of the velocity, ${v_{t}} =1$ ms$^{-1}$, is imposed at inlet and outlet surfaces, as shown in Figure \ref{Figure17}, while the secondary velocity, normal to the axial direction, is set null, ${v_{n}=0}$. Pressure $p=0$ is prescribed at the outlet, and the non-slip boundary condition, $\mathbf{v}=0$, is imposed at the pipe's walls. Initial conditions ${v_t}=1$ ms$^{-1}$, ${v_{n}=0}$ and $p=0$, for velocity and pressure, are considered.

\begin{figure}[!htb]
\begin{center}
\includegraphics[width=0.70\textwidth]{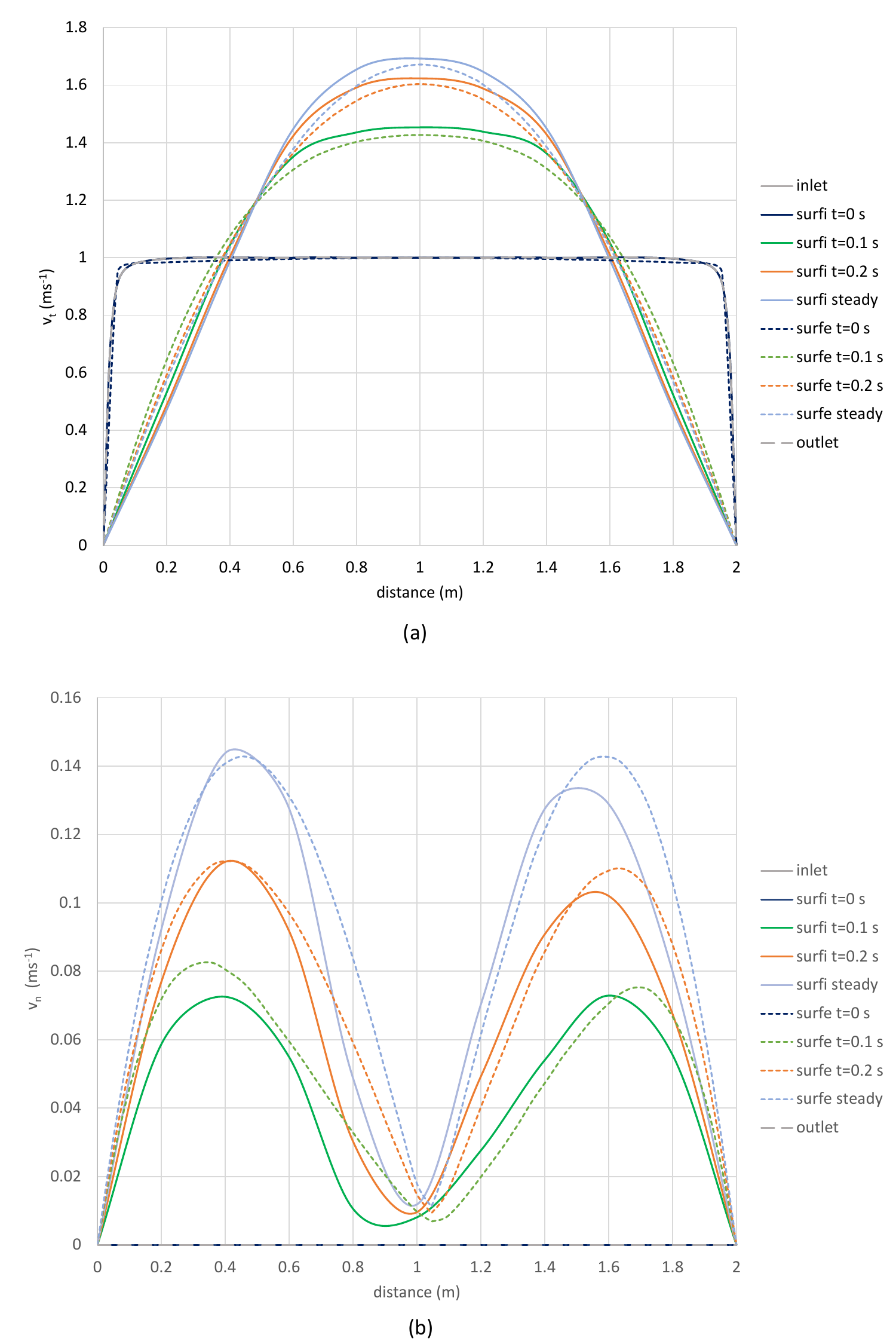}
\end{center}
\caption{Helical flow under fixed boundary conditions: (a) axial component of velocity; (b) secondary velocity (normal to the axial component) at inlet, outlet and control surfaces \textit{i} and \textit{e}. Same curve color indicates same time; same curve marker indicates same surface.}\label{Figure18}
\end{figure}

\begin{figure}[!htb]
\begin{center}
\includegraphics[width=0.70\textwidth]{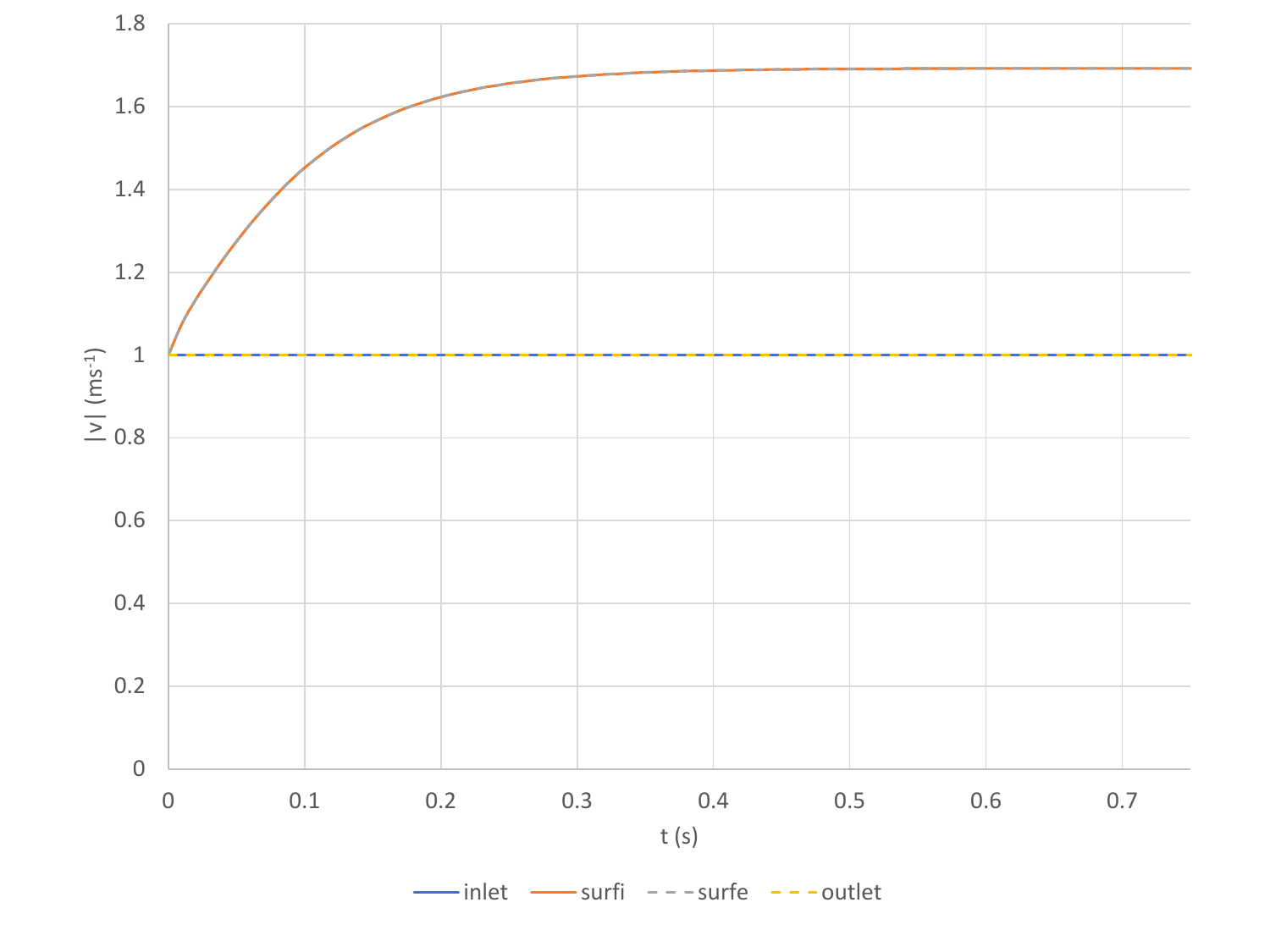}
\end{center}
\caption{Helical flow under fixed boundary conditions: evolution of the modulus of velocity at the pipe's central axis for inlet, outlet and control surfaces \textit{i} and \textit{e}. }\label{Figure19}
\end{figure}

\begin{figure}[!htb]
\begin{center}
\includegraphics[width=0.70\textwidth]{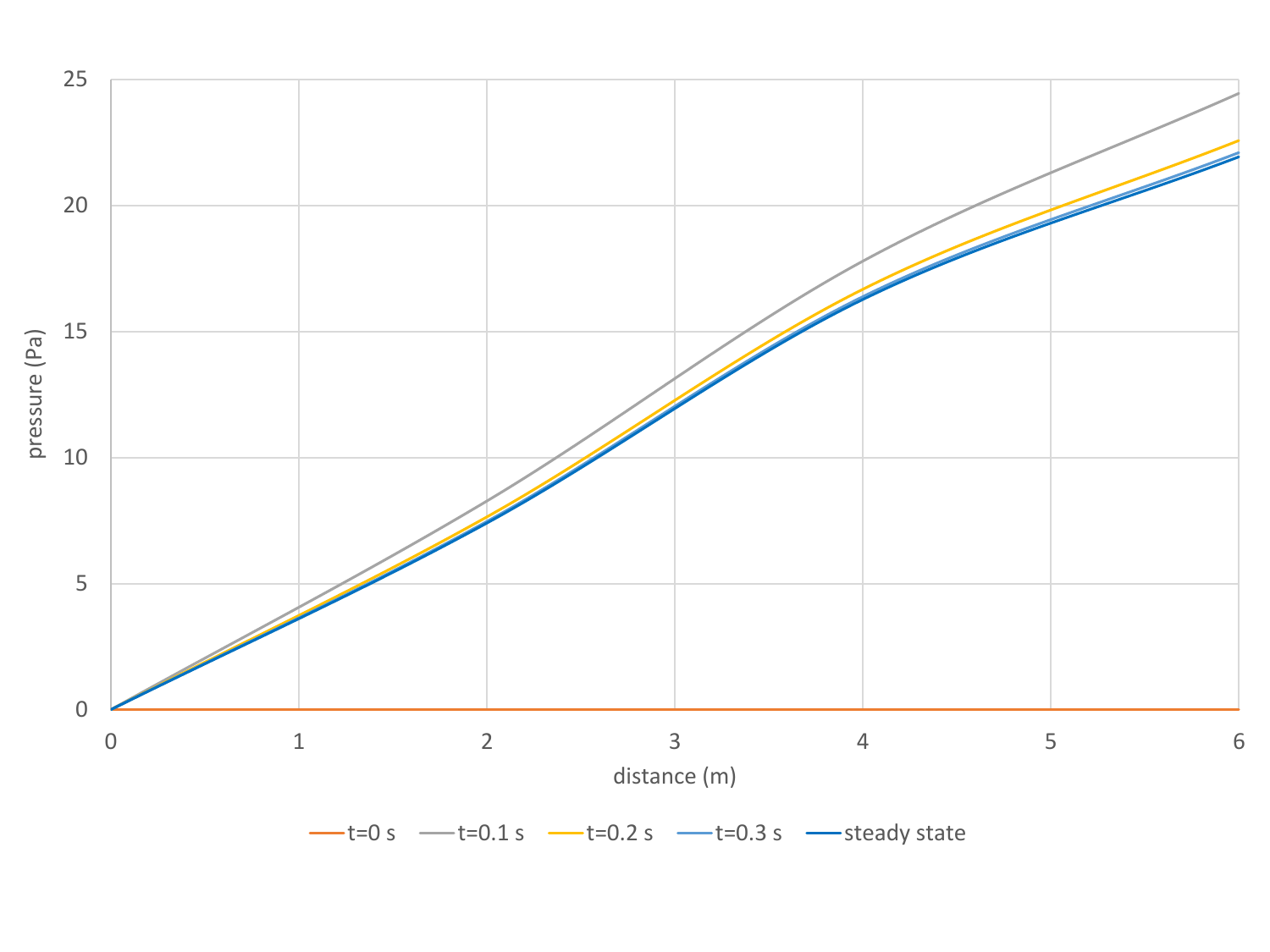}
\end{center}
\caption{Helical flow under fixed boundary conditions: pressure variation along the pipe's central axis. Distance is measured starting from the outlet section and projected over the helix' revolution axis.}\label{Figure20}
\end{figure}

The fluid is considered to be newtonian and incompressible, with density $\rho = 1$ kg m$^{-3}$ and dynamic viscosity $\mu = 0.5$ Pa s. The time step used in the computation is $\Delta t = 0.5 \cdot 10^{-4}$ s. The results obtained for the evolution in time of the axial and secondary velocities inside the pipe for the four sections considered are shown in Figures \ref{Figure18}a and \ref{Figure18}b, respectively. Figure \ref{Figure19} depicts the evolution of the velocity modulus in time at the center on the four considered sections. The pressure evolution along the pipe's central axis is depicted in Figure \ref{Figure20}. Note here that distance in the horizontal axis of graph \ref{Figure20} is measured over the helix' revolution axis, and therefore the values obtained for the pressure along the pipe's central axis are projected in that direction. It is worth mentioning here that the pressure variation along the pipe is not linear as it was in the Poiseuille flow (Fig. \ref{Figure12}), due to the presence of the sectional secondary flow, as a consequence of the centrifugal force. 

As observed in Figs.\ref{Figure18} and \ref{Figure19}, the solution for the velocity inside the volume depends of the section considered, proving that the flow inside the pipe is \textit{not} fully developed. Variations in the secondary flow pattern cause deviations from the linear variation of the pressure, as observed in Fig. \ref{Figure20}. It is worth noting here that the observed discrepancies in the solutions for velocity and pressure at the different sections are a direct result of the non-fully developed nature of the flow, and by no means a numerical artifact.  

\begin{figure}[!htb]
\begin{center}
\includegraphics[width=0.70\textwidth]{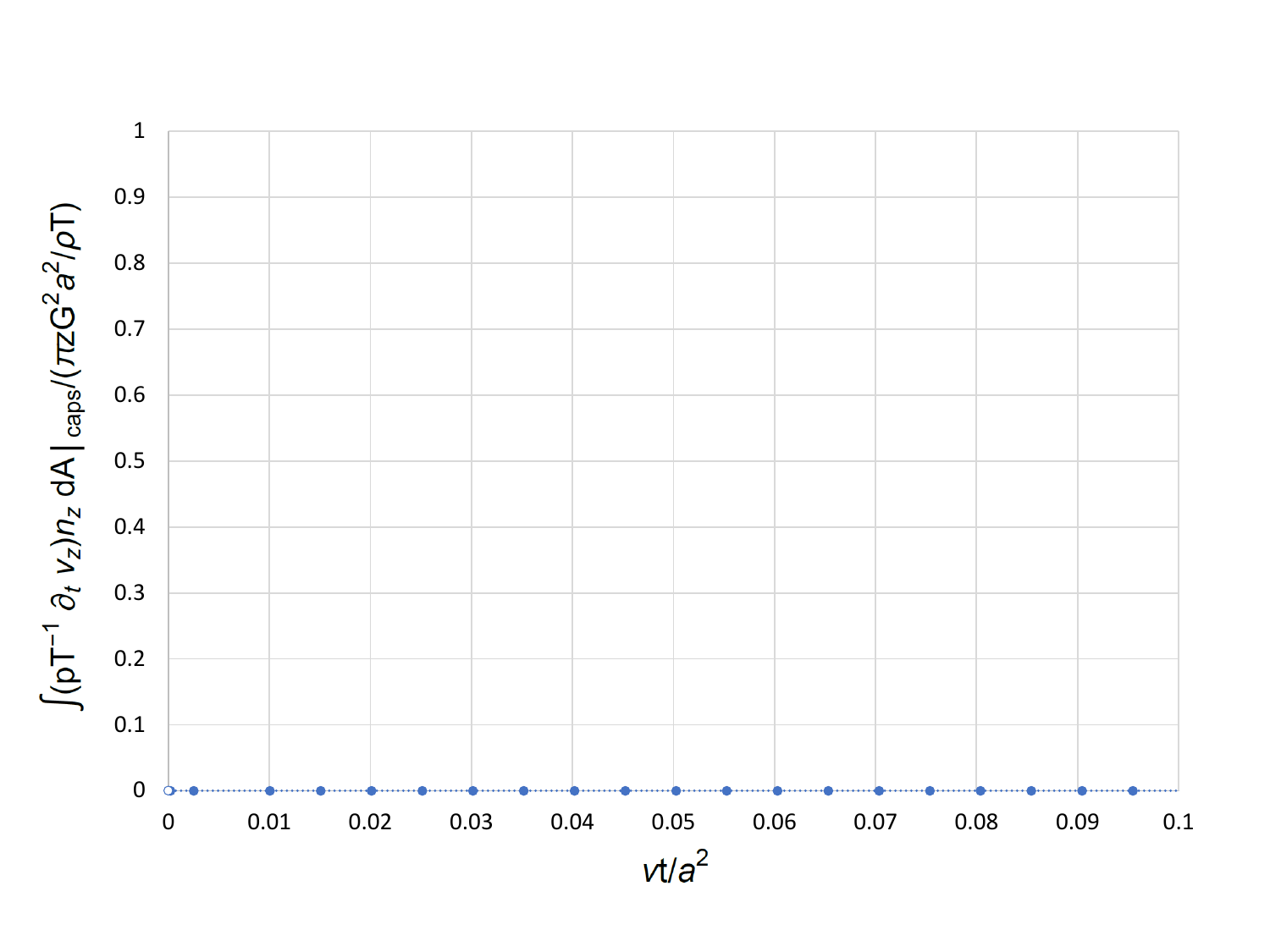}
\end{center}
\caption{Helical flow under fixed boundary conditions: Null surface contribution - Normalized solution of Eq. (\ref{surfaceFEM}).}\label{Figure21}
\end{figure}

\begin{figure}[!htb]
\begin{center}
\includegraphics[width=0.70\textwidth]{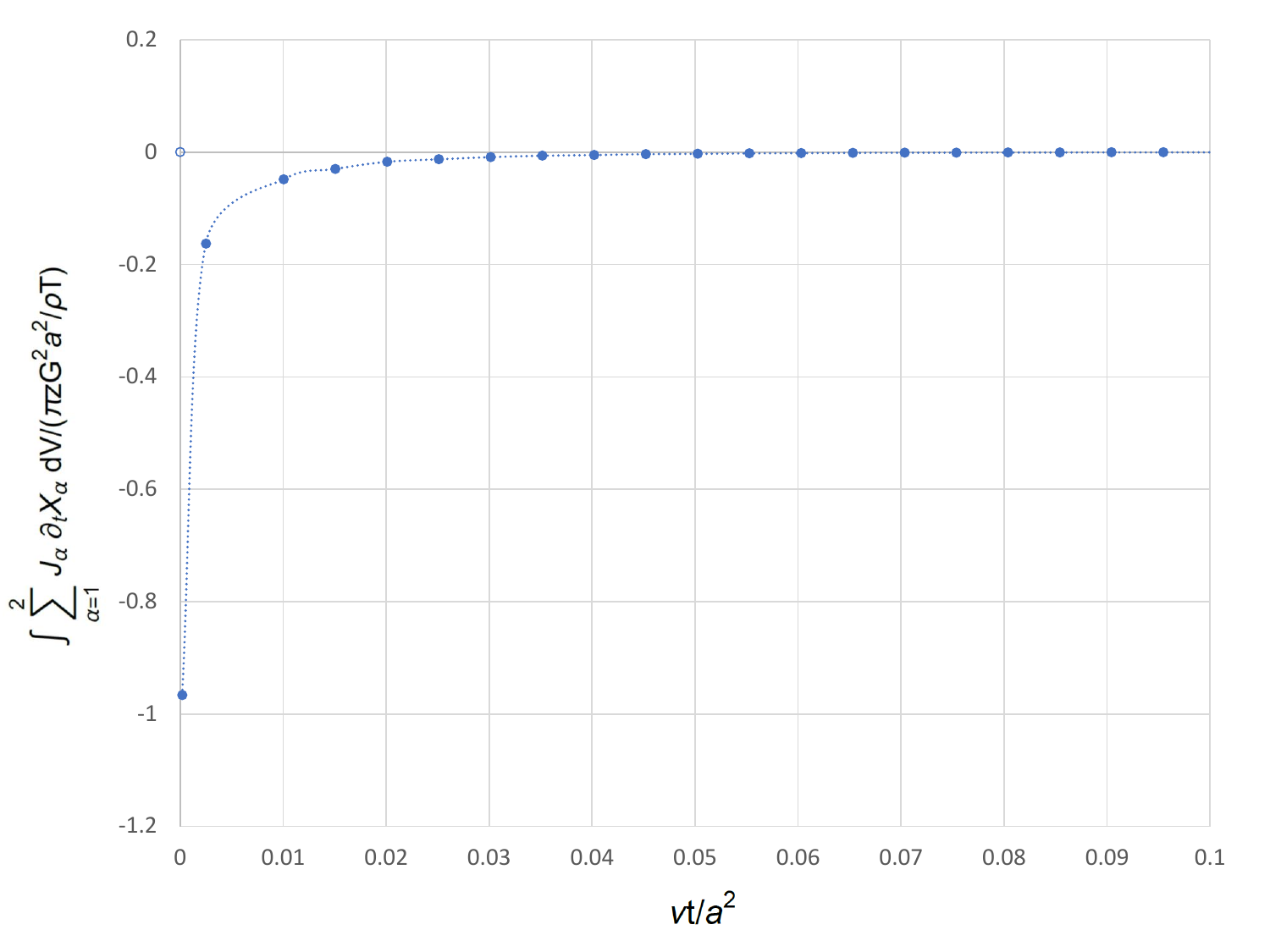}
\end{center}
\caption{Helical flow under fixed boundary conditions: Negative volume contribution - Normalized solution of Eq. (\ref{surfaceFEM}).}\label{Figure22}
\end{figure}

\begin{figure}[!htb]
\begin{center}
\includegraphics[width=0.70\textwidth]{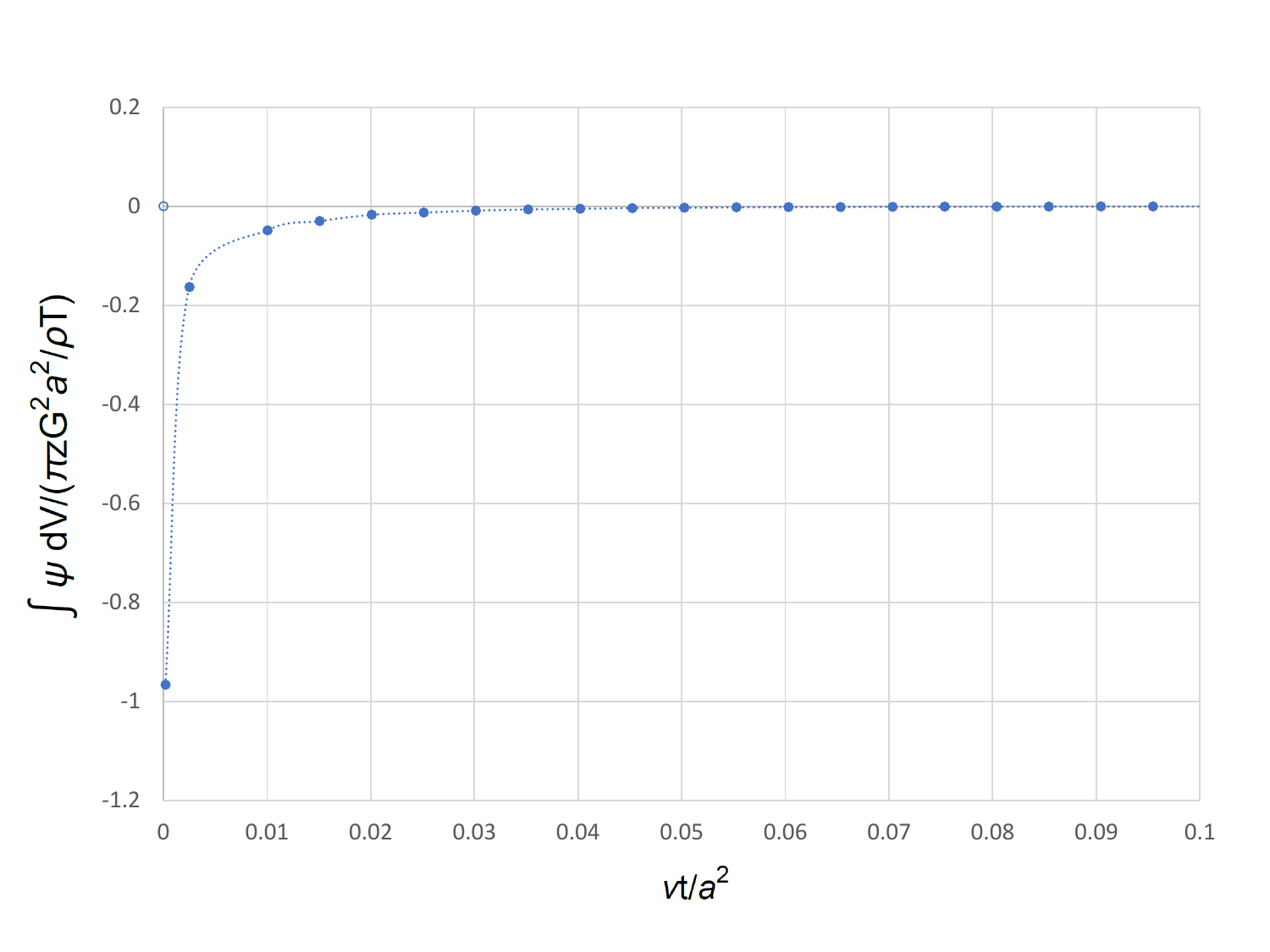}
\end{center}
\caption{Helical flow under fixed boundary conditions: Full negative contribution - Normalized solution of Eq. (\ref{surfaceFEM}).}\label{Figure23}
\end{figure}

Evaluation of the extended evolution criterion is carried out once velocity and pressure fields are computed inside the domain, as a function of time. Figures \ref{Figure21}, \ref{Figure22} and \ref{Figure23} show the normalized curves obtained from computing Eqs.(\ref{surfaceFEM}), (\ref{volumeFEM}) and (\ref{EGEC-FEM}) respectively. 

These results confirm numerically that the inequality in Eq. (\ref{weakGEC}) \textit{does} still holds for the helical flow problem. On the contrary to the starting flow problem for the Poiseuille flow, this result coincides with the GEC as firstly stated by Glansdorff and Prigogine, since the evolution of this system involves \textit{fixed} boundary conditions \cite{GP1964,GPbook}. Thus, a perfect accordance is found in this case between the GEC and the extended evolution criterion (EGEC) presented herein.


\subsection{\label{sec:helical_t} Helical flow subjected to time-dependent boundary conditions}

Finally, we test the extended evolution criterion for a curved pipe subjected to time-dependent boundary conditions. The setup of the problem is the same as explained in Section \ref{sec:helical-fixed}, and summarized in Figure \ref{Figure16}a. 

\begin{figure}[!htb]
\begin{center}
\includegraphics[width=0.70\textwidth]{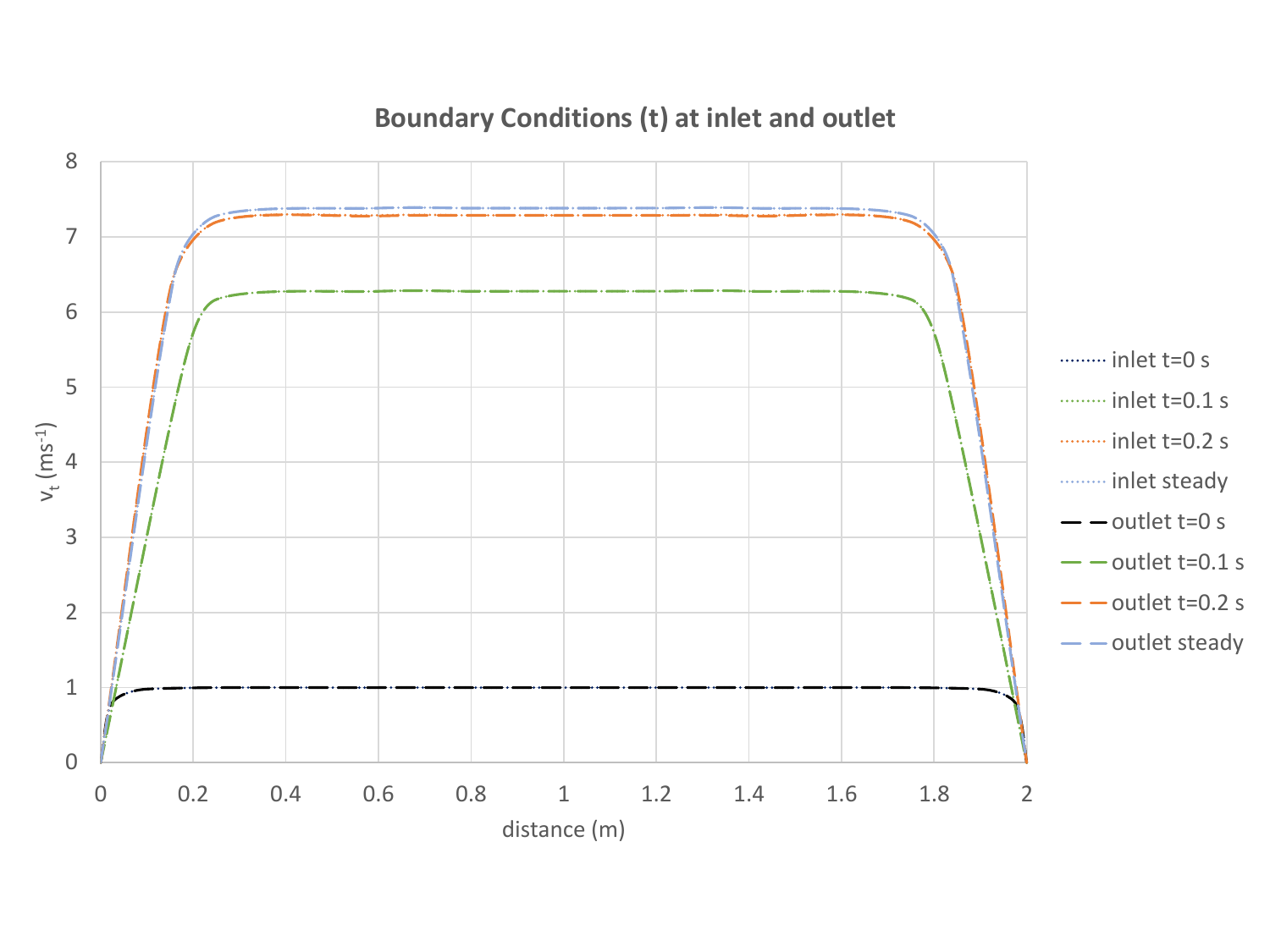}
\end{center}
\caption{Helical flow subjected to time-dependent boundary conditions: Time-dependent boundary conditions for the axial velocity at inlet and outlet. The curve color indicates the time while the curve marker indicates the surface;}\label{Figure24}
\end{figure}

Same 3D computational mesh of 48314 linear tetrahedra and 9175 nodes as explained in Section \ref{sec:helical-fixed} is used for the computation (see Figure \ref{Figure16}b). The boundary condition depicted in Figure (\ref{Figure24}) for the axial component of the velocity, ${v_{t}}$, is imposed at inlet and outlet surfaces, while the secondary velocity, normal to the axial direction, is set null, ${v_{n}=0}$. Pressure $p=0$ is prescribed at the outlet, and the non-slip boundary condition, $\mathbf{v}=0$, is imposed at the pipe's walls. Initial conditions ${v_t}=1$ ms$^{-1}$, ${v_{n}=0}$ and $p=0$, for velocity and pressure, are considered.

The fluid is considered newtonian and incompressible, with same density, $\rho = 1$ kg m$^{-3}$, and dynamic viscosity, $\mu = 0.5$ Pa s, as in former sections. The time step used in the computation is again $\Delta t = 0.5 \cdot 10^{-4}$ s. 

\begin{figure}[!htb]
\begin{center}
\includegraphics[width=0.70\textwidth]{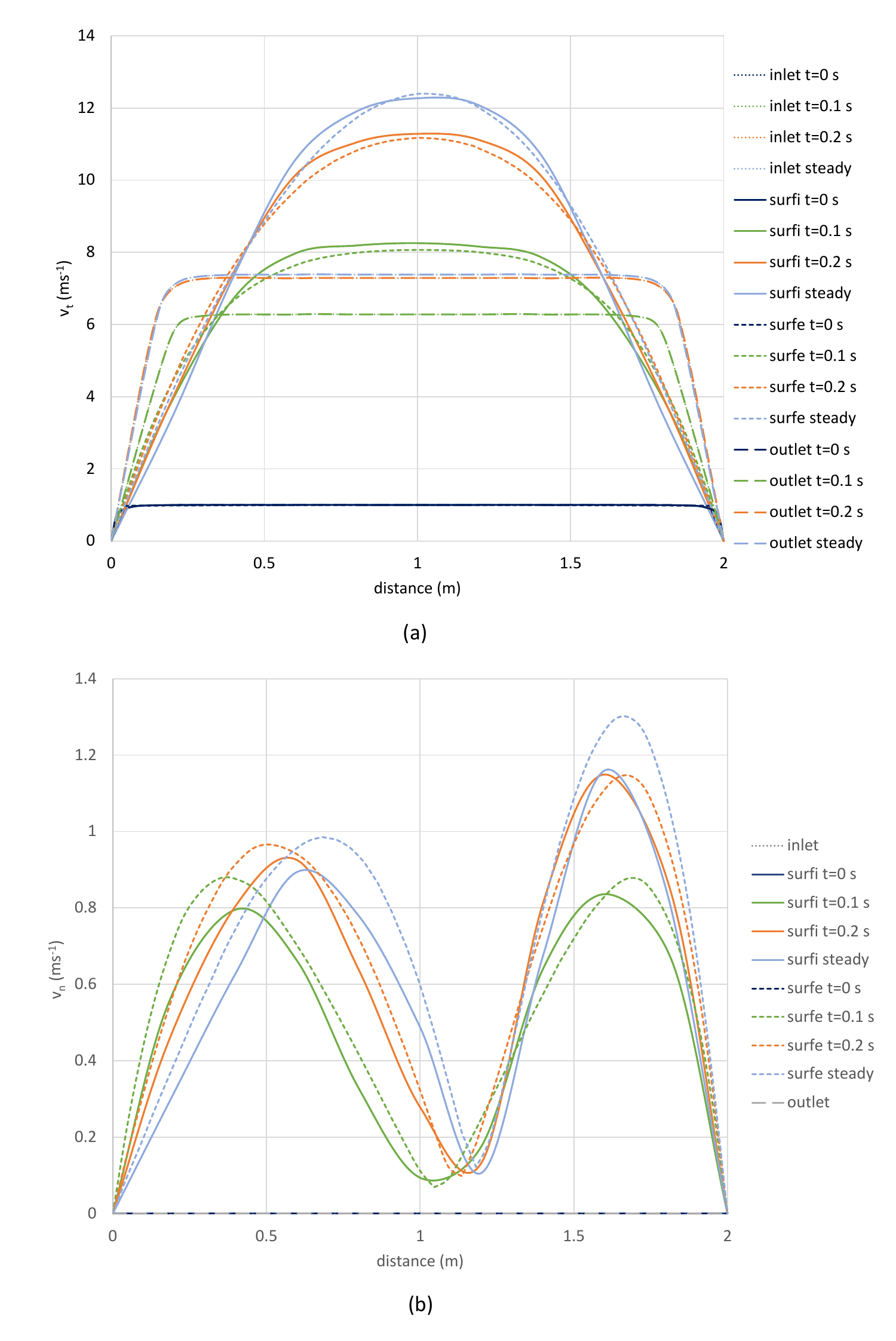}
\end{center}
\caption{Helical flow subjected to time-dependent boundary conditions: (a) axial component of velocity; (b) secondary velocity (normal to the axial component) at inlet, outlet and control surfaces \textit{i} and \textit{e}. Same curve color is used for the same time while same curve marker refers to the same surface.}\label{Figure25}
\end{figure}

\begin{figure}[!htb]
\begin{center}
\includegraphics[width=0.70\textwidth]{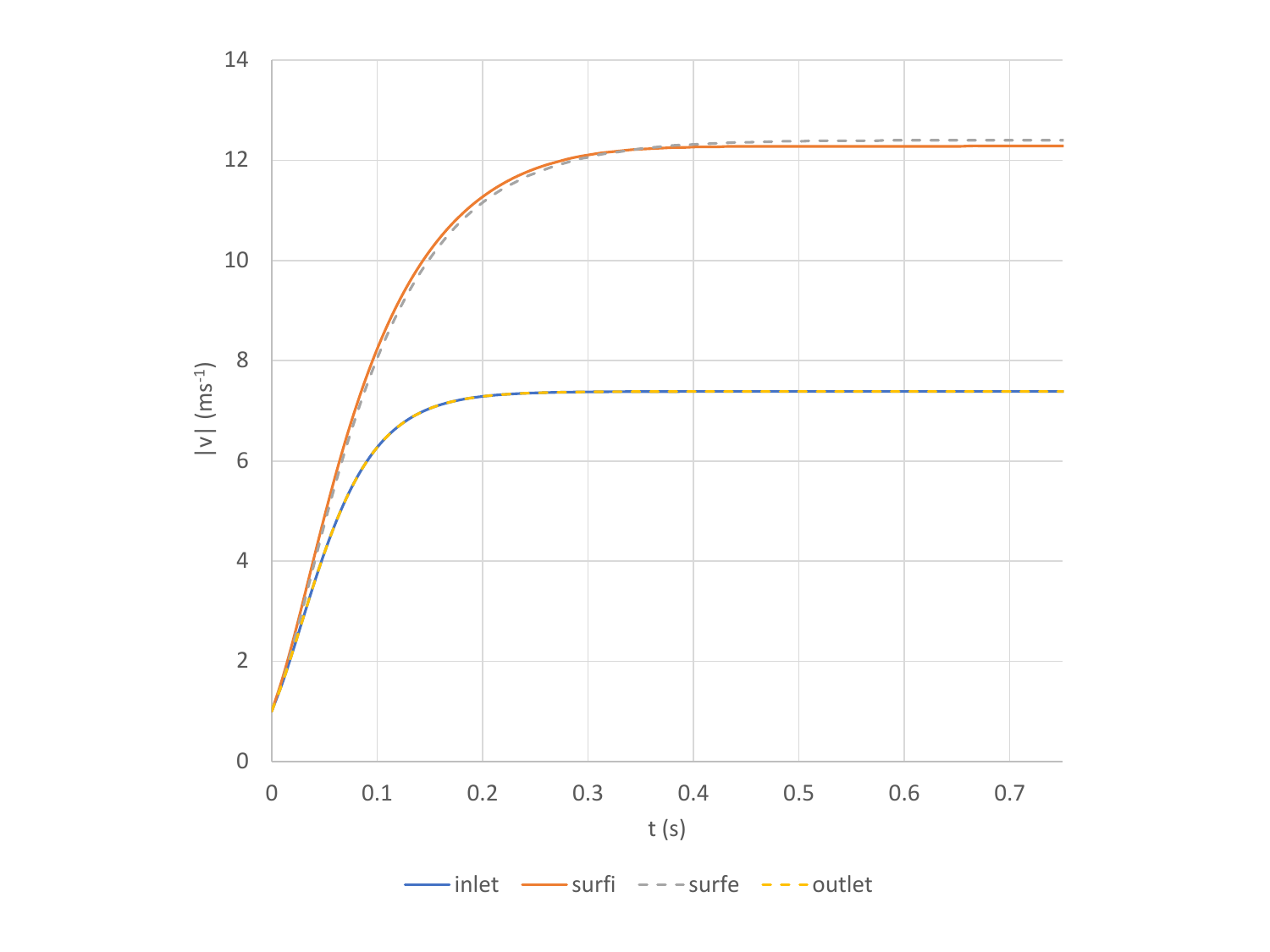}
\end{center}
\caption{Helical flow subjected to time-dependent boundary conditions: evolution of the modulus of velocity at the pipe's central axis for inlet, outlet and control surfaces \textit{i} and \textit{e}. }\label{Figure26}
\end{figure}

\begin{figure}[!htb]
\begin{center}
\includegraphics[width=0.70\textwidth]{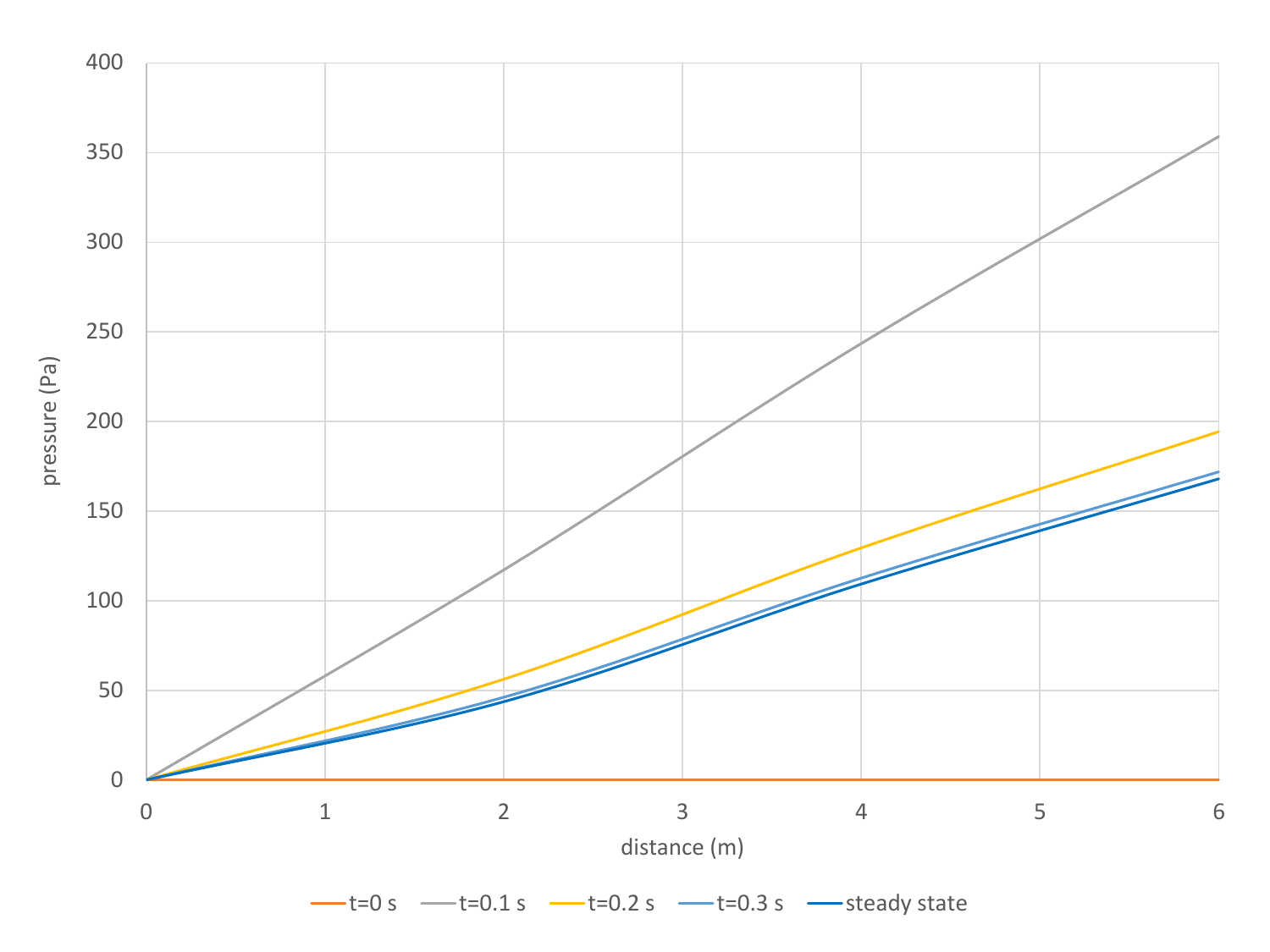}
\end{center}
\caption{Helical flow subjected to time-dependent boundary conditions: pressure gradient along the pipe's central axis. Distance is measured starting from the outlet section and projected over the helix' revolution axis.}\label{Figure27}
\end{figure}

The results obtained for the evolution in time of the axial and secondary velocities inside the pipe for the four sections considered are shown in Figures \ref{Figure25}a and \ref{Figure25}b, respectively. Figure \ref{Figure26} depicts the evolution of the velocity modulus in time at the center on the four considered sections. The pressure evolution along the pipe's central axis measured over the helix' revolution axis is depicted in Figure \ref{Figure27}. 

As observed in Figs. \ref{Figure25} and \ref{Figure26}, the solution for the velocity inside the volume depends on the section considered, showing again that the flow is \textit{not} fully developed inside the pipe. Comparing the secondary flow for the helical pipe with fixed boundary conditions (Fig. \ref{Figure18}b) and time-varying boundary conditions (Fig. \ref{Figure25}b) we observe a time-dependent right-shift for the location at which the minimum occurs (i.e. the position of the Dean vortex). This is due to the fact that while the flow regime (i.e. the Reynolds number) is constant for the helical pipe with fixed boundary conditions, this is not the case for the time-dependent boundary conditions where the flow regime depends on time, and so does the secondary flow pattern. Nevertheless, the interested reader can find the details on the dynamics of the secondary flow in helical pipes and the formation and evolution of Dean vortices in \cite{HHPoF} and references therein.


\begin{figure}[!htb]
\begin{center}
\includegraphics[width=0.70\textwidth]{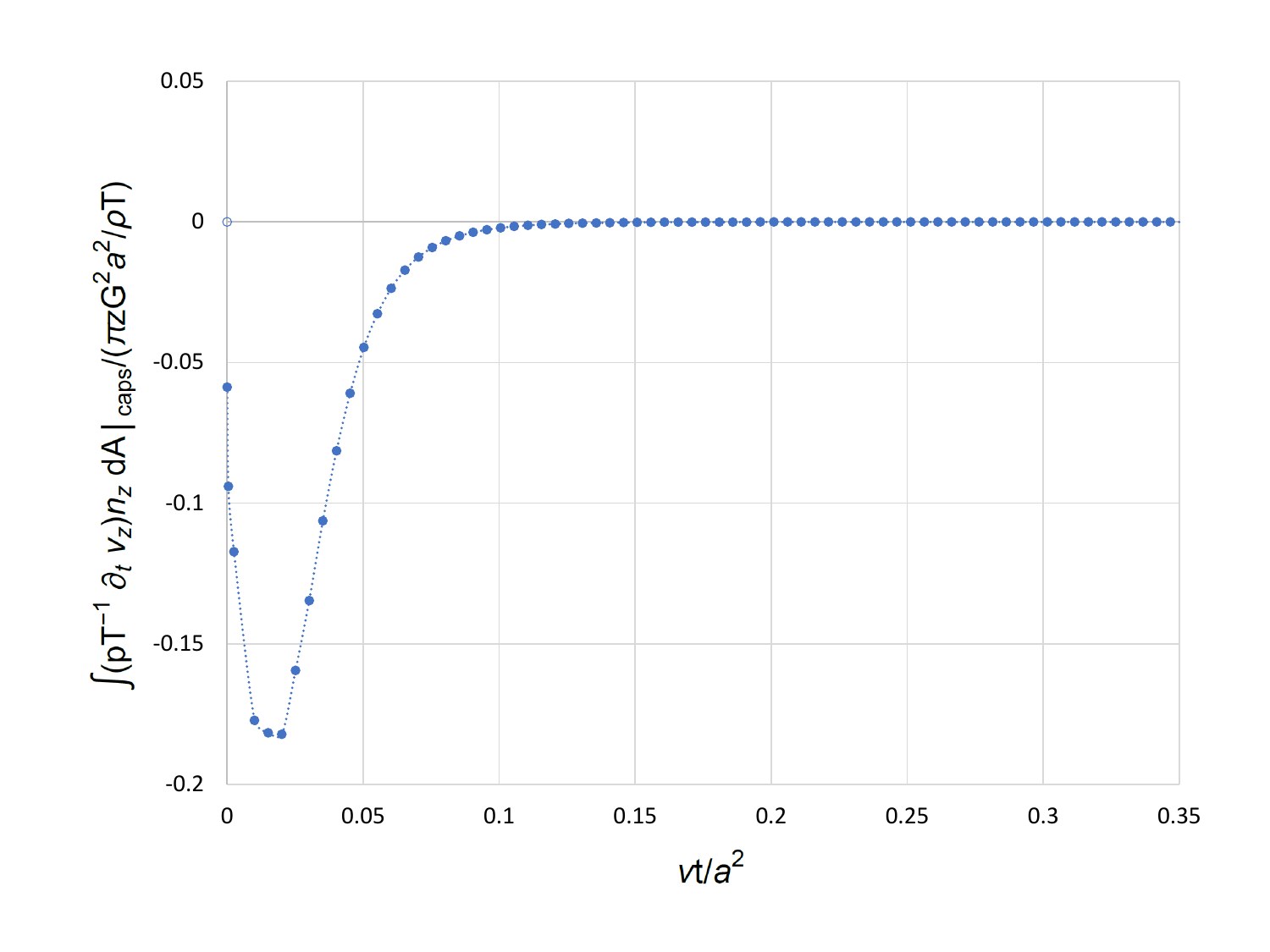}
\end{center}
\caption{Helical flow subjected to time-dependent boundary conditions: Negative surface contribution - Normalized solution of Eq. (\ref{surfaceFEM}). }\label{Figure28}
\end{figure}

\begin{figure}[!htb]
\begin{center}
\includegraphics[width=0.70\textwidth]{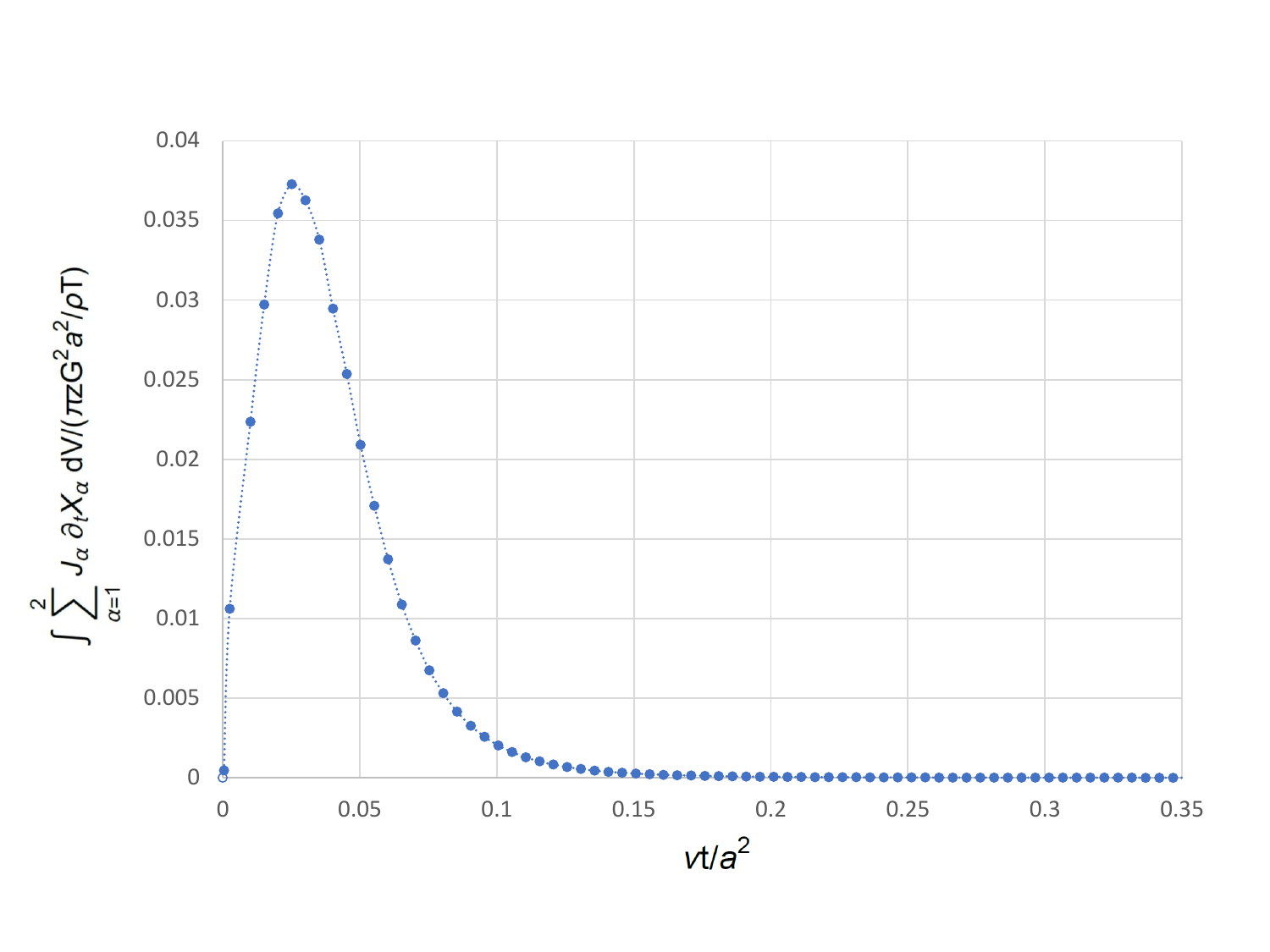}
\end{center}
\caption{Helical flow subjected to time-dependent boundary conditions: Positive volume contribution - Normalized solution of Eq. (\ref{surfaceFEM}).}\label{Figure29}
\end{figure}

\begin{figure}[!htb]
\begin{center}
\includegraphics[width=0.70\textwidth]{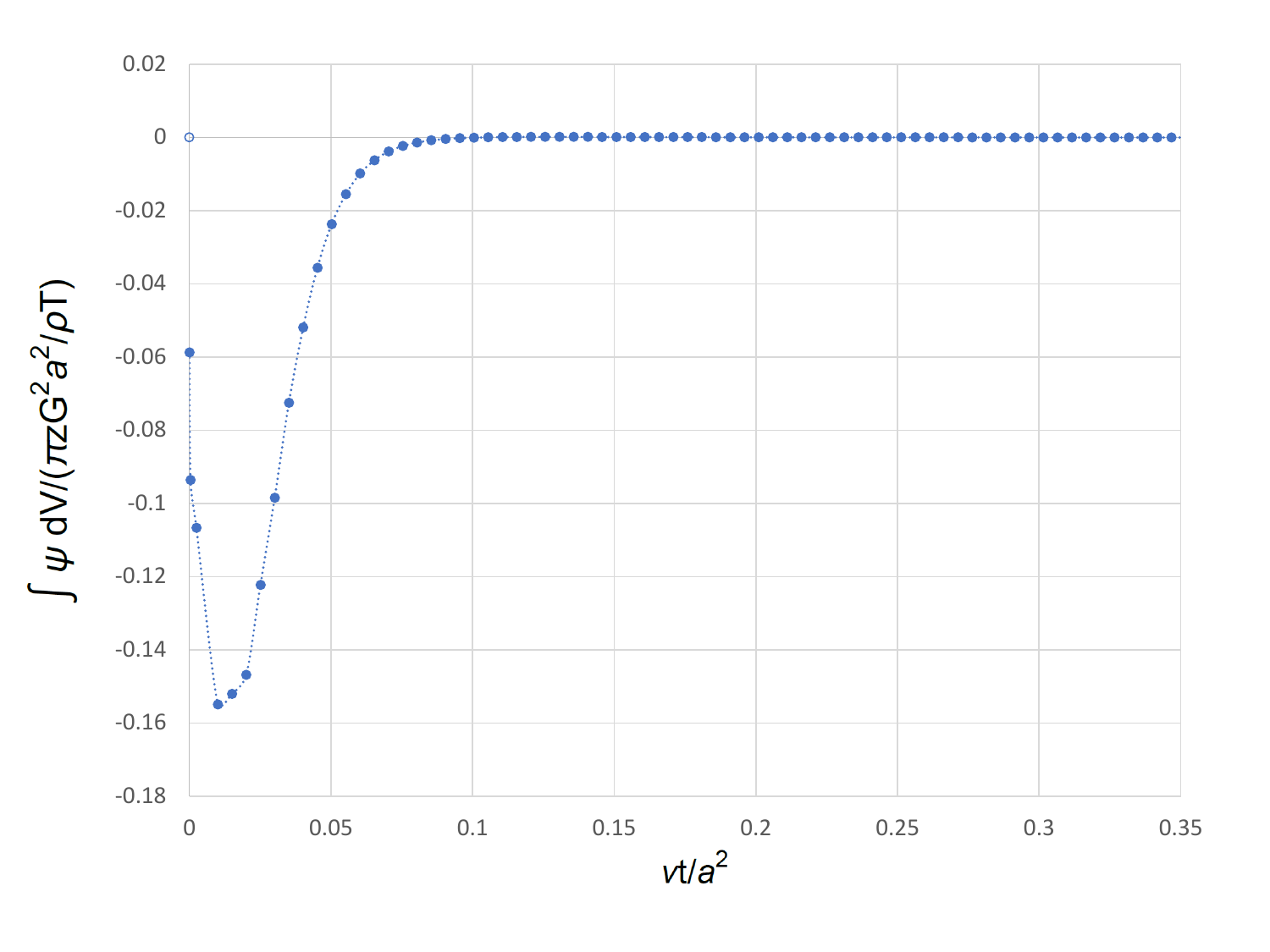}
\end{center}
\caption{Helical flow subjected to time-dependent boundary conditions: Full negative contribution - Normalized solution of Eq. (\ref{surfaceFEM}).}\label{Figure30}
\end{figure}

Evaluation of the extended evolution criterion is carried out using the computed velocity and pressure fields. Figures \ref{Figure28}, \ref{Figure29} and \ref{Figure30} show the normalized curves obtained from computing Eqs.(\ref{surfaceFEM}), (\ref{volumeFEM}) and (\ref{EGEC-FEM}) respectively. 

These results confirm once more that the inequality in Eq. (\ref{weakGEC}) \textit{does} hold for the helical flow subjected to time-dependent boundary conditions. However, this result does not coincide with the GEC. Therefore, it has been proven that  the extended evolution criterion given in (\ref{weakGEC}) is required to properly account for the time-dependent boundary problem in the most general case.

\clearpage
\section{\label{sec:disc} Discussion}

We validated analytically and numerically, an extended version of the General Evolution Criterion (EGEC), which allows for arbitrary time-dependent boundary conditions. 
The ultimate goal is to acquire a deeper understanding of the thermodynamic and mechanical evolution of open systems, in the context of viscous fluid flows.

This extended evolution criterion has been specialized to cover time-dependent convective processes. Within this framework, the velocity field on the system's boundary is allowed to exhibit any time-dependent behavior. The resultant inequality places constraints on the production of entropy within the system, emphasizing the role 
of the surface or boundary contribution. Boundary conditions can be chosen, and these correspond to the thermodynamic and mechanical state of the external environment interacting with the system. Thus, the state of the external environment exerts a direct influence on the internal thermodynamic and mechanical evolution of the system.  External influences, as accounted for by the boundary conditions, play a pivotal role in shaping how the system's thermodynamic forces evolve over time. Inspection of the 
inequality Eq.(\ref{weakGEC}) of EGEC leads to the following assertions. Namely, if the surface contribution is null (time-independent boundary conditions) then the volume contribution is necessarily negative definite (GEC).  If the surface term is positive, the volume contribution must be sufficiently negative to counteract the former to obey the inequality. However, if the surface term is negative, then the sign of the volume term can be either positive or negative. The sign of the volume contribution tells us whether the 
entropy production of the system tends to increase (positive) or else decrease (negative). 

For illustrative purposes, we considered specific examples. By carefully selecting the fluid velocity on the entrance and exit bounding surfaces of both straight and curved pipes, we can either maximize or minimize the internal fluid viscous entropy production within the bulk volume. These examples underscore the practical implications of this extended GEC (EGEC), and demonstrate the potential for controlling and optimizing the behavior of open systems in fluid dynamics through judicious choice of boundary conditions.  

A comprehensive examination of the velocity field and viscous entropy production within a Poiseuille-type flow under time-dependent boundary conditions was carried out through both mathematical analysis and numerical simulations. Our findings confirm an important observation: the time-dependent viscous entropy consistently undergoes an incremental transition from zero to a definitive positive value as we vary the pipe radius and maintaining a null value solely along the central axis of the pipe.

We used the established analytical solution for the initial flow problem to assess the surface and volume integrals featured in the extended evolution inequality. Our analysis revealed distinct trends: the volume contribution yields a positive value across all time instances and asymptotically approaches zero from above. This result provides a clear counterexample to the General Evolution Criterion (GEC), which can hold only for time-independent boundary conditions.

Conversely, the surface contribution exhibits a negative value across all time intervals and converges asymptotically towards zero from below. Notably, we have demonstrated a precise mathematical counterbalance between the surface contribution and the principal positive term within the volume contribution. This cancellation results in a finite remainder for the volume contribution, which remains negative definite and tends towards zero from below as the system approaches its non-equilibrium steady state (NESS). This underscores the obeyance of the initial flow problem to the EGEC. We emphasize that this validation hinges critically upon the intricate interplay between surface and volume effects.

After successfull validation of our numerical approximation using the Poiseuille starting flow problem, we applied it to a more intricate scenario: the not fully developed flow within a non-trivial geometric configuration, characterized by a curved and torsional pipe in order to demonstrate the general validity of the EGEC.
We assessed EGEC for a helical pipe flow, considering two contrasting scenarios: (i) fixed boundary conditions and (ii) time-dependent boundary conditions. Under fixed boundary conditions, our calculations of the velocity components at the inlet, outlet, and two internal control surfaces within the helical conduit corroborated that the flow remains not fully developed. Additionally, the secondary flow velocity analysis revealed the presence of a single recirculating vortex. These findings are in accord with our prior investigations into flow within curved pipes with torsion \cite{HHPoF}. In this context, the surface integral computation yielded a null outcome, consistent with fixed boundary conditions, while the volume integral contributed a negative value that asymptotically approaches zero from below as the system approaches its NESS. Consequently, in this case, the GEC holds.
However, when we imposed time-dependent boundary conditions, the surface integral exhibited time-dependent negative values, while the volume integral becomes positive definite. Importantly, only the joint effect of these two contributions obeyed the extended inequality defined in Eq. (\ref{weakGEC}), thereby demonstrating that helical flow with time-dependent boundary conditions conforms to the dictates of the EGEC, but not those of the GEC.

These outcomes hold substantial implications, not only for the domain of fluid mechanics but also in broader contexts, including the comprehension of the mecano-thermodynamic evolution  of living organisms, open dissipative systems continuously exchanging energy and matter with their surrounding environment. To this end, in future work we aim to include
heat and matter transport phenomena as well as chemical reactions. 

\section*{Acknowledgements}
This research has been funded by grant No. PID2020-116846GB-C22 by
the Spanish Ministry of Science and Innovation/State Agency of
Research MCIN/AEI/10.13039/501100011033 and by "ERDF A way of making
Europe". I.H. expresses her gratitude for these years of scientific collaboration with the late D.H., who is mourned by family, friends, and colleagues. 

\section*{Data availability} The data that support the
findings of this study are available from the authors of this study
upon reasonable request.

\section*{Declaration of Interests} The authors report no conflict of interest.

\appendix

\section{\label{sec:matrix} Matrix notation}

Here we put the surface and volume contributions to the evolution
criterion in matrix notation.

Define:
\begin{eqnarray}\label{matrixnotation}
\bm{v} &=& (v_x\,v_y\,v_z)\\
\bm{n} &=& (n_x\,n_y\,n_z)\\
\bm{\nabla}\bm{v} &=& [\partial_i v_j] = \left(
                                           \begin{array}{ccc}
                                             \partial_x v_x & \partial_x v_y & \partial_x v_z \\
                                             \partial_y v_x & \partial_y v_y & \partial_y v_z \\
                                             \partial_z v_x & \partial_z v_y & \partial_z v_z \\
                                           \end{array}
                                         \right)\\
\mathbf{D}= d_{ij} &=& \mu \Big(\frac{\partial v_i}{\partial x_j} +
\frac{\partial v_j}{\partial x_i}\Big) = \mu \Big([\partial_j v_i] +
[\partial_i v_j] \Big) = \mu \big(
\bm{\nabla}\bm{v} +(\bm{\nabla}\bm{v})^T \big)
\end{eqnarray}
Then for the \textsf{surface terms} in Eq.(\ref{surface}) we have,
where superindex $T$ stands for transpose:
\begin{eqnarray}
\big(p T^{-1} (\partial_t v_j) \big)n_j &=& p T^{-1}\, (\partial_t \bm{v}) \, \bm{n}^T\\
-\big(d_{ij} T^{-1} \partial_t v_i \big)n_j &=& -T^{-1} \mu\,(\partial_t \bm{v})
\, \big(\bm{\nabla}\bm{v}+(\bm{\nabla}\bm{v})^T\big)\,\bm{n}^T = -T^{-1} (\partial_t \bm{v}) \, \mathbf{D} \,
\bm{n}^T
.
\end{eqnarray}
Next, for the \textsf{volume terms} in Eq.(\ref{GECreduction2}) we
have $t_r$ stands for the \textsf{trace}:
\begin{eqnarray}
+T^{-1} d_{ij}\, \partial_t(v_{i,j}) &=& +T^{-1} \mu \, t_r [
\big(\bm{\nabla}\bm{v}+(\bm{\nabla}\bm{v})^T \big)\, (\partial_t \bm{\nabla} \bm{v})]=+T^{-1} \, t_r [
\mathbf{D} \, (\partial_t \bm{\nabla} \bm{v})]\\
\rho T^{-1} v_{j} v_{k,j}\,
\partial_t(v_k) &=& \rho T^{-1} v_{j} \,[\partial_j v_k]
\partial_t(v_k) =
\rho T^{-1} \bm{v} \, \bm{\nabla}\bm{v} \, (\partial_t \bm{v})^T
\end{eqnarray}

\end{document}